\documentclass[12pt]{article}

\usepackage{scicite}

\usepackage{times}

\usepackage{amsmath}
\usepackage{amsfonts}
\usepackage{amssymb}
\usepackage{graphicx}
\usepackage{ulem}

\usepackage{bm,color,amsmath,amssymb,latexsym,graphicx,psfrag,accents,float}
\usepackage{amscd,dsfont,wasysym,mathrsfs,mathtools}
\usepackage{multirow}
\usepackage{dcolumn}
\usepackage{xcolor}
\usepackage{comment}

\usepackage{booktabs} 

\usepackage{lineno}

\definecolor{AAcolor}{rgb}{0.7,0.1,0.4}

\newcommand{\f}[2]{\frac{#1}{#2}}

\newcommand\as{\;\;\;\;}

\newcommand{\la}[1]{\label{#1}}

\newcommand{\q}[1]{Eq.\ (\ref{#1})}

\newcommand{\var}{\varepsilon}
\newcommand{\bk}{\boldsymbol{k}}
\newcommand{\bp}{\boldsymbol{p}}
\newcommand{\imp}{\;\;\Rightarrow\;\;}

\title{Temperature dependence of quantum oscillations from non-parabolic dispersions}

\author
{Chunyu Guo,${}^{1\ast\dagger}$ A. Alexandradinata,${}^{2,3\ast\dagger}$ Carsten Putzke,${}^{1}$ Amelia Estry,${}^{1}$ Teng \\ Tu,${}^{4}$ Nitesh Kumar,${}^{5}$ Feng-Ren Fan,${}^{5}$ Shengnan Zhang,${}^{6,7}$ Quansheng Wu,${}^{6,7}$ \\Oleg V. Yazyev,${}^{6,7}$ Kent R. Shirer,${}^{5}$ Maja D. Bachmann,${}^{5,8}$ Hailin Peng,${}^{4}$ Eric\\ D. Bauer,${}^{9}$ Filip Ronning,${}^{9}$ Yan Sun,${}^{5}$ Chandra Shekhar,${}^{5}$  Claudia Felser,${}^{5}$ \\ Philip J. W. Moll,${}^{1\dagger}$\\
	\normalsize{${}^{1}$Laboratory of Quantum Materials (QMAT), Institute of Materials (IMX),}\\ \normalsize{\'{E}cole Polytechnique F\'{e}d\'{e}rale de Lausanne (EPFL), CH-1015 Lausanne, Switzerland}\\
	\normalsize{${}^{2}$Institute for Condensed Matter Theory, University of Illinois at Urbana-Champaign,}\\ \normalsize{ Urbana, Illinois 61801, USA}\\
	\normalsize{${}^{3}$Department of Physics, University of Illinois at Urbana-Champaign, }\\ \normalsize{Urbana, Illinois 61801, USA}\\
	\normalsize{${}^{4}$Center for Nanochemistry, Beijing National Laboratory for Molecular Sciences (BNLMS),}\\ \normalsize{College of Chemistry and Molecular Engineering, Peking
		University, Beijing 100871, China.}\\
	\normalsize{${}^{5}$Max Planck Institute for Chemical Physics of Solids, 01187 Dresden, Germany}\\
	\normalsize{${}^{6}$Chair of Computational Condensed Matter Physics (C3MP), Institute of Physics (IPHYS),}\\ \normalsize{\'{E}cole Polytechnique F\'{e}d\'{e}rale de Lausanne (EPFL), CH-1015 Lausanne, Switzerland}\\
	\normalsize{${}^{7}$National Centre for Computational Design and Discovery of Novel Materials MARVEL, }\\ \normalsize{\'{E}cole Polytechnique F\'{e}d\'{e}rale de Lausanne (EPFL), CH-1015 Lausanne, Switzerland}\\
	\normalsize{${}^{8}$School of Physics and Astronomy, University of St Andrews, St Andrews KY16 9SS, UK}\\
	\normalsize{${}^{9}$Los Alamos National Laboratory, Los Alamos, New Mexico 87545, USA}\\
	\\
	\normalsize{$^\ast$These authors contributed equally to this work.}\\
	\normalsize{$^\dagger$Corresponding authors: chunyu.guo@epfl.ch(C.G); aalexan7@illinois.edu(A.A.);}
	\\
	\normalsize{philip.moll@epfl.ch(P.J.W.M.)}\\}

\date{}

\begin{document}

	\baselineskip24pt

	\maketitle

	\section*{Abstract}

	The phase offset of quantum oscillations is commonly used to experimentally diagnose topologically non-trivial Fermi surfaces. This methodology, however, is inconclusive for spin-orbit-coupled metals where $\pi$-phase-shifts can also arise from non-topological origins. Here, we show that the linear dispersion in topological metals leads to a $T^2$-{temperature correction} to the oscillation frequency that is absent for parabolic dispersions. We confirm this effect experimentally in the Dirac semi-metal Cd$_3$As$_2$ and the multiband Dirac metal LaRhIn$_5$. Both materials match a tuning-parameter-free theoretical prediction, emphasizing their unified origin. For topologically trivial Bi$_2$O$_2$Se, no frequency shift associated to linear bands is observed as expected. However, the $\pi$-phase shift in Bi$_2$O$_2$Se would lead to a false positive in a Landau-fan plot analysis. Our frequency-focused methodology does not require any input from ab-initio calculations, and hence is promising for identifying correlated topological materials.

	\section*{Introduction}
	
	The discovery of topological semimetals promises an avenue to study novel materials that host quasiparticles that mimick relativistic Dirac and Weyl fermions in high-energy physics\cite{zhijun_na3bi,zhijun_cd3as2,DiracAndWeyl_Armitage_RMP,bradlyn2017topological}. They host bands {that touch at points or lines in momentum space}; such degeneracies are typically associated to closed Fermi surfaces with topologically-robust Berry phases. While initially considered a rare occurrence, recent ab-initio programs\cite{bradlyn2017topological,TopoCatalogue,XGW_topoSearch,TopoCatalogue2} {have predicted topological band degeneracies in a {sixth} of all non-magnetic materials in {the crystal database}\cite{TopoCatalogue}.  }
	
	Magnetic quantum oscillations\cite{shoenberg2009magnetic} promise to play a key role in experimentally confirming these predictions. It is widely believed that a $\pi$-phase shift in quantum oscillations is intepretable as a $\pi$ Berry phase, and therefore a smoking-gun confirmation {of a topological semimetal}. Such phase analysis, often carried out with a `Landau-fan plot', ignores other non-geometric phase shifts, and forgets that the Berry phase is not quantized to a integer multiple of $\pi$ for many symmetry classes of (semi)metals\cite{phiB_Aris_PRX,100page}. For this reason, an unambiguous topological diagnosis is generally impossible for the 3D Dirac semimetals\cite{zhijun_na3bi,zhijun_cd3as2,DiracAndWeyl_Armitage_RMP} and low-symmetry 3D Weyl semimetals\cite{phiB_Aris_PRX}. 
	
	For these cases, we present a new identification method based on the temperature ($T$) dependence of the oscillation frequency ($F$), finding a characteristic $T^2$ contribution that is uniquely attributed to the linear energy-momentum dispersion of Dirac, Weyl and multifold fermions\cite{Bradlyn_newfermions}. This $T^2$ contribution is a 3D, higher-degeneracy generalization of an effect predicted by K\"ubbersbusch and Fritz\cite{KupperbuschFritz} for 2D Dirac materials. To the best of our knowledge, this effect has never been experimentally studied in graphene, yet we will show that it is easy to see in 3D Weyl/Dirac materials. Our strategy applies to candidate topological Fermi pockets which are small compared to the Brillouin-zone volume; small pockets are accurately described by $\boldsymbol{k}\cdot \boldsymbol{p}$ Hamiltonians that retain only the leading-order term -- giving a parabolic dispersion for the Schr\"odinger-type fermion, and a linear dispersion for the Dirac-type fermion. These two cases are distinguishable by the {energy-derivative} of the cyclotron mass $m_c$ (Fig. 1). While $\partial m_c/\partial E=0$ for a parabolic dispersion, for a linear dispersion $E(k)=\pm \sqrt{(v_x\hbar k_x)^2+(v_y\hbar k_y)^2}$, the particular energy dependence of the Fermi-surface area $S$ yields a non-zero energy derivative of the cyclotron mass:
	\begin{equation}
	S=\f{\pi}{\hbar^2} \f{E_F^2}{v_xv_y} \imp m_c=\f{\hbar^2}{2\pi}\bigg|\f{\partial S}{\partial E}\bigg|=\f{|E_F|}{v_xv_y} \imp \f1{m_c}\bigg|\f{\partial m_c}{\partial E}\bigg|=\f1{|E_F|},\la{dmdE}
	\end{equation}
	with $E_F$ the Fermi energy measured from the Dirac point and $v_j$ the Fermi velocity.

	To experimentally determine $\partial m_c/\partial E$, we exploit the fact that quantum oscillations probe the band structure over an energy window ($\sim k_BT$) around the Fermi level, due to the thermal broadening of the Fermi-Dirac distribution function. As lighter-mass particles experience less thermal damping than heavier particles, the effective frequency renormalizes towards lighter orbits as $T$ increases. This effect is absent for parabolic bands because the effective mass is energy-independent. However, for a Dirac-type Fermi surface, the frequency decreases with increasing $T$ because the effective mass is smaller closer to the node, as illustrated in Fig. 1(a). This temperature-renormalization of $F$ applies generally to linear dispersions near band degeneracies, including Dirac and Weyl degeneracies, as well as higher-fold degeneracies associated to Fermi pockets with higher Chern numbers (so-called `multifold fermions')\cite{Bradlyn_newfermions}.
	
	\section*{Results}
	\subsection*{General methodology}
	To quantify this frequency shift, {it is useful to view the Lifshitz-Kosevich formula\cite{lifshitz_kosevich_jetp,Roth_SOCandZeeman} as an asymptotic expansion in powers of $k_BT/E_F$ (the degeneracy parameter of Fermi gases). Odd powers of $T$ modify the oscillation amplitude, with the first odd power giving the well-known thermal damping factor; in contrast, even powers of $T$ modify the frequency and phase.} At elevated temperatures where {$2\pi^2 k_BT$ is large or comparable} to the cyclotron energy, we derive in Sec. II A (of the Supplementary Material) a $T^2$-correction to the oscillation frequency:
	\begin{equation}
	F_0(\mu)\rightarrow {F}(\mu,T)=F_0(\mu)+\Delta F^{top}(T),\as \Delta F^{top}(T):=- \f{\pi^2}{4}\f{(k_BT)^2}{\beta} \bigg|\f{\partial (\log m_c)}{\partial E}\bigg|,\la{correctF}
	\end{equation}
	with $\mu$ the chemical potential and $\beta:=e\hbar/2m_c$ the effective Bohr magneton. The correction $\Delta F^{top}$ is our main theoretical result, and applies to {any closed Fermi surface originating from an {arbitrary} energy-momentum dispersion, whether linear, quadratic, cubic or beyond}. While $\Delta F^{top}$ vanishes for parabolic bands, it is finite for linear bands as  $|\partial (\log m_c)/{\partial E}|= 1/|E_F|$, according to \q{dmdE}, and we hence refer to $\Delta F^{top}$ as a {topological correction} to $F$.

	Next we consider other mechanisms for temperature dependence of the frequency. A distinct $T^2$ correction arises from the temperature dependence of the chemical potential [$\mu(T)$] at fixed particle density, as is well-known in the Sommerfeld theory of metals\cite{AshMer} [Fig. 1(b)]. The sum of Sommerfeld and topological corrections is expressed in:
	\begin{equation}
	F(\mu,T) = F_0(E_F)- \Theta \frac{(\pi k_BT)^2}{\beta^2F_0(E_F)}+O(T^4),
	\la{Fparabolicpaper}
	\end{equation}
	with $\Theta$ a dimensionless coefficient. 
	In high-carrier-density metals with both small and large pockets, the Sommerfeld correction is negligible. The frequency shift of a small pocket is reduced by a factor $|E_F|/E_{bw}\ll 1$, with $E_F$ the Fermi energy of the small pocket measured from band extremum/degeneracy, and $E_{bw}$ the typical bandwith. Thus we expect that $\Theta$ is dominated by the topological correction, giving $\Theta\approx 0$ in the parabolic case, and $\Theta \approx 1/16$ in the case of a linear dispersion. For single-frequency, low-carrier-density semimetals, the Sommerfeld correction is of the same order of magnitude as the topological correction, giving $\Theta=1/48$ for parabolic bands and $\Theta=5/48$ for linear bands ({see Supplementary Sec.II B for extended calculation of Sommerfeld correction}). The two scenarios show that the observation of a $T^2$ correction alone is not conclusive of nontrivial topology. Rather, conclusiveness comes from
	the following experimental consistency check: since $F_0(E_F)$, $m_c$ and additionally the frequency shift at elevated temperatures all can be experimentally determined, by applying \q{Fparabolicpaper} one obtains an experimental value of $\Theta$ which should consistently equal the conditional values that our theory predicts.
	
	In principle, the entropic contribution of electrons gives an additional $T^2$ correction owing to the band-structure modification by thermal expansion\cite{barron1980thermal}. However, this correction to $F$ is typically of parts in $10^4$ up to the highest temperature that quantum oscillations are observable\cite{shoenberg2009magnetic}. {Frequency shifts with a $T^4$ dependence have been observed for a few, non-magnetic metals; these shifts were attributed to the {lattice} contribution to thermal expansion\cite{PhysRev.95.1421,PhysRev.151.484}, as well as the electron-phonon coupling\cite{lonzarich1983temperature} (see Supplementary Sec. II A). Because of their distinct power law ($T^4$) compared to the topological and Sommerfeld corrections ($T^2$), lattice and electron-phonon effects would in principle be easy to detect and analyze separately.}
	
	Next we discuss the identification of band topology. Our approach senses the linearity of bands and hence the topological character needs to be inferred. As the argument is based on a $\boldsymbol{k}\cdot \boldsymbol{p}$ expansion, it is only applicable to small Fermi surfaces that are much smaller than the Brillouin zone. As topological materials of practical relevance generally host small Fermi pockets, this criterion is typically fulfilled. Even when linear bands with a small Fermi pocket are detected, a question remains about distinguishing massive from massless Dirac materials. Conservatively speaking, one can never completely rule out the possibility that any hypothesized Dirac fermion has a tiny mass $m_D$ which leads to a weakly-nonlinear energy dispersion
	$E(k)=\pm \sqrt{(v_x\hbar k_x)^2+(v_y\hbar k_y)^2+m_D^2}$ and  $\Theta=(1/16)|E_F|(|E_F|+2|m_D|)/(|E_F|+|m_D|)^2.$ {The experimental uncertainty in $\Theta$ can be used to set an upper bound on $|m_D|$. If the dispersion at the Fermi level is experimentally indistinguishable from a linear one, a hypothetical gap at the node is necessarily much smaller than the chemical potential (measured from the node), and has negligible influence on the low-energy excitations. For example, graphene with a typical chemical potential ($\sim meV$)  is well described by massless Dirac fermions, despite the existence of a spin-orbit-induced gap ($\sim \mu e V$)\cite{SOCGap_Graphite}. }
	
	High temperatures are natural opponents of quantum oscillations, because discontinuous changes in the occupation of Landau levels are smoothened out by the Fermi-Dirac distribution, {as illustrated in the top panel of Fig. 2}. The characteristic $T^2$ dependence for $F$ is observable at a temperature scale $T^*$ where $2\pi^2k_BT^*$ is comparable to the cyclotron energy, while for $T\gg T^*$ oscillations are exponentially suppressed by thermal damping\cite{shoenberg2009magnetic}. The {optimal} temperature window for observing $F(T)$ is determined approximately by plotting the product (of the frequency shift and the amplitude) as a function of $T$, as in Fig. 2. Since $T^*$ is inversely proportional to the effective mass $m_c$, the requirement for elevated temperatures does not preclude their observation even when quasiparticles masses are heavy, it simply reduces the optimal temperature window to lower values. For example, for $m_c$ that is ten times the free-electron mass, Fig. 2 predicts the optimal temperature window to be between $0.1 - 1$ K. Our method hence is expected to apply to strongly interacting topological materials with strong mass renormalization. {Detailed discussion on applicability to heavy-fermion materials can be found in supplement Sec. II C.}
	\subsection*{Detection and analysis of temperature dependence of quantum oscillation frequency}
	
	Experimentally, these predictions turn out to be readily observable. Three distinct materials were analyzed (Fig. 3): (i) Cd$_3$As$_2$ is a well-studied prototypical Dirac semimetal with a  time-reversal-related pair of Dirac-type Fermi pockets and no other pockets ($F \approx$ 43.7 T)\cite{CdAs_review}. (ii) Bi$_2$O$_2$Se is a topologically trivial semimetal with a single, small electron pocket centered at $\Gamma$ ($F \approx$ 33.3 T)\cite{BiOSe}. (iii) LaRhIn$_5$ is a large-carrier-density, multi-band metal that hosts Brillouin-zone-sized pockets\cite{LaCe115_dHvA_jpsj} in addition to a very small pocket ($F \approx$ 6.9 T) that is inconsistent with conventional Schr\"odinger-like behavior\cite{La115_goodrich_PRL}. In their pioneering work, Mikitik and Sharlai proposed that the small pocket encloses a Dirac nodal line\cite{La115_Miktik_PRL}, based on an assumption that the spin-orbit coupling is perturbatively weak. However our first-principles calculation  ({detailed in the Supplementary Sec. II D and Sec. IV}) suggest this assumption to be unjustified, leaving the topology of the small pocket still in question. 
	
	Crystalline microbars of (i-iii) for four-terminal resistivity measurements were prepared by Focused Ion Beam machining\cite{moll2018focused}. These microbars feature optimized geometries for longitudinal transport and provide high signal amplitudes even in the highly conductive materials studied here. All samples show pronounced quantum oscillations of the longitudinal magnetoresistance, with a single small frequency in the low-field regime (Fig. 3). We obtain $F_0(E_F)$ from extrapolating the temperature-dependent oscillation frequency to zero temperature, and $m_c$ from the temperature dependence of the amplitude. {For each material, the temperature dependence of the frequency is obtained from fitting the entire experimental dataset [measured at different temperatures and fields (see Fig. 4)] to a single Lifshitz-Kosevich formula.\cite{phiB_Aris_PRX} The fitting is done by a standard least squares regression method using the non-linear model fitting function provided by Mathematica, as detailed in Supplementary Sec. III A and B.}
	

	Both Cd$_3$As$_2$ and Bi$_2$O$_2$Se are low-carrier-density materials in which the Sommerfeld correction applies, and accordingly a clear frequency shift, $\Delta F(T)$, is observed in both of them (Fig. 4). It is evident from the raw data that Cd$_3$As$_2$ exhibits a stronger frequency shift compared to the trivial Bi$_2$O$_2$Se. For both cases, $\Delta F(T)$ falls directly onto the theoretical predictions $\Theta {(\pi k_BT)^2}/{\beta^2F_0(E_F)}$ from \q{Fparabolicpaper} for the trivial and topological case respectively. For Cd$_3$As$_2$, $\Theta =5/48$, as predicted for a Dirac semimetal {with only two time-reversal-related Fermi pockets;  for Bi$_2$O$_2$Se,  $\Theta=1/48$ is consistent with a conventional semimetal with a single Fermi pocket}. We emphasize that the coefficient of $T^2$ has {no tuning parameter} as $F_0(E_F)$ and $m_c$ are fixed by measurement results, i.e., theory fixes $\Theta$ to take on different rational values depending on whether the pocket is Schr\"odinger- or Dirac-type. 
	
	In comparison, LaRhIn$_5$ is a high-carrier-density metal, and hence the measured $\Delta F(T)$ (for three devices) directly matches the prediction of $\Theta=1/16$, a purely topological correction. This confirms the small 7 T pocket of LaRhIn$_5$ is Dirac-type, with a chemical potential that is pinned by other large coexisting pockets. 
	
	The universal topological aspect of these distinct compounds becomes evident after subtracting the Sommerfeld correction {described by Supplementary Eq. (10) and (11)} from the experimental $\Delta F$ for Bi$_2$O$_2$Se and Cd$_3$As$_2$. Remarkably, Cd$_3$As$_2$ and all three devices of LaRhIn$_5$ collapse on the same red line in Fig. 4(c) despite their highly different band structures and microscopic details, highlighting the common topological origin of $\Delta F^{top}$ and its insensitivity to material-specific details. Instructively, a quantum-oscillation phase analysis of Bi$_2$O$_2$Se and Cd$_3$As$_2$ by the Landau-fan-diagram method uncovers a $\pi$-phase shift in both of them [see Fig. 4{(d)}], despite their clearly distinct topology, exemplifying the faults of this method.

	These results can be further quantitatively strengthened by the self-consistency of $|E_F|$ computed by two different ways. From $\bk\cdot\bp$ theory, it can be expressed in terms of the standard Lifshitz-Kosevich parameters: $|E_F|  =2e\hbar F_0/m_c$ for the linearized Dirac pocket, and $|E_F|  =e\hbar F_0/m_c$ for the quadratic Schr\"odinger pocket. On the other hand, $|E_F|$ can also be determined via $\Delta F(T)$: in the Dirac case,  $|E_F| = |\f{\partial E}{\partial (\log m_c)}|$ [cf.\ \q{dmdE}] is deducible from the topological correction, while in the Schr\"odinger case $|E_F|$  is deducible\cite{AshMer} from the Sommerfeld correction to $F$. For the model correctly describing the topology of the pocket, both estimates for $|E_F|$ should be consistent, and indeed this is what Table 1 shows. This allows for a simple self-consistency check. By analyzing a measured temperature-dependent quantum oscillation frequency $F(T)$ within both the Dirac/Weyl and Schr\"odinger framework, the match of both $|E_F|$ signals the correct $\bk\cdot\bp$ model.

	

	\section*{Discussions}
	To conclude, our experimental methodology allows to diagnose the linear dispersion of small, topological Fermi pockets, as demonstrated by our three case studies. Despite their microscopic differences, all Dirac/Weyl/multifold fermions have a linear dispersion close to the nodal degeneracy. The linear dispersion is directly sensed by the temperature dependence of the oscillation frequency when the Fermi level is also close to the node. It is also capable of identifying multi-Weyl fermions protected by crystallographic rotational symmetry, e.g., the dispersion of a double-Weyl (triple-Weyl) fermion is quadratic (cubic) in two momentum directions and linear in the third direction\cite{MultiWeyl}, hence they would be identifiable by measuring $F(T)$ at various field orientations. Our methodology is applicable independent of the magnitude of the Zeeman splitting of Landau levels; this magnitude would affect the relative amplitudes of higher harmonics of quantum oscillations\cite{phiB_Aris_PRX} but not their frequency.
	
	{Because the energy scale for band inversion tends to be small compared to the bandwidth, topological Fermi pockets are often small compared to the Brillouin zone. It is not impossible for highly-inverted materials that the dispersion of larger topological Fermi pockets acquires substantial nonlinear corrections, in which case our methodology becomes less useful for topological diagnosis. It is also less straightforward to subtract the Sommerfeld correction (from the frequency shift) in materials where a topological Fermi pocket coexists with a non-topological pocket of comparable size; here, supplementing our methodology with a first-principles calculation may be useful to isolate the topological frequency shift.}
	
	It will be interesting to apply our method to investigate the topology of Fermi-liquid materials with extreme interaction-driven mass enhancement (e.g., heavy-fermion materials), with the caveat that the optimal temperature for measuring $F(T)$ is inversely proportional to the effective mass. The presence of Dirac fermions in LaRhIn$_5$ suggests that the intimately-related Ce(Co, Rh, Ir)In$_5$ are prime candidates for such efforts. The Ce family has a similar band structure to LaRhIn$_5$, but are more prone to correlation-driven instabilities such as unconventional superconductivity\cite{Ce115_JDT_review}. A different class of correlated topological metals include the Kondo-Weyl semimetals\cite{lai2018weyl,WeylKondoSurf}, in which itinerant electronic bands hybridize with localized (4f/5f)-states leading to strongly-renormalized Weyl dispersions. 
	
	As the field matures towards strongly-interacting topological matter, it leaves the comfort zone in which ab-initio-predicted topological band structures can straightforwardly be confirmed by angle-resolved photoemission spectroscopy. The new conceptual and experimental challenges -- arising from competing, near-degenerate ground states and unconventional superconductivity -- call for new approaches to detect low-energy excitations and assess their topological character. Extending the framework of quantum oscillations to topologically nontrivial Fermi surfaces, as presented here, will play a major role in this development.

	\bibliographystyle{Nature}

	\clearpage
	
	\section*{Methods}
	
	\noindent \textbf{Microstructure fabrication.}
	Micro-devices of all materials are fabricated with a FEI Helios Plasma FIB using Xe-ions. Thin slab (lamella) was dig out from a single crystalline material, which was later transferred and glued down to a sapphire substrate. The transferred lamella was later patterned to the desired geometry with the Plasma FIB. \\
	
	\noindent \textbf{Band structure calculations.}
	To check for the robustness of band structure results against choices of functionals in the calculation, the calculations were performed via two independent methods which are both detailedly described in the Supplementary Sec. IV A.\\
	
	\noindent \textbf{Data Availability.} 
	Data that support the findings of this study is deposited to Zenodo with the access link: https://doi.org/10.5281/zenodo.5482689.\\
	
	\noindent \textbf{Code Availability.} 
	Mathematica code used for the Lifshitz-Kosevich fit of temperature-dependent quantum oscillations can be
	found via the access link:  https://doi.org/10.5281/zenodo.5482689.\\

	\section*{Acknowledgements} 
	\noindent \textbf{Funding.} A.A. was supported initially by the Yale Postdoctoral Prize Fellowship, and subsequently by the Gordon and Betty Moore Foundation EPiQS Initiative through Grant No. GBMF 4305 and GBMF 8691 at the University of Illinois. S.Z, Q.W. and O. V. Y. acknowledge support by NCCR Marvel and the Swiss National Supercomputing Centre (CSCS) under Projects No. s832 and No. s1008. M.D.B. acknowledges studentship funding from EPSRC under grant no EP/L015110/1. E.D.B. and F.R. were supported by the U.S. DOE, Basic Energy Sciences, Division of Materials Sciences and Engineering. C.P. and P.J.W.M. acknowledge funding by the European Research Council (ERC) under the European Union's Horizon 2020 research and innovation programme ("MiTopMat" - grant agreement No. 715730). This project was funded by the Swiss National Science Foundation (Grants  No. PP00P2\_176789).\\
	\noindent \textbf{Author Contributions.} Crystals were synthesized and characterized by  F.R., E.D.B. (LaRhIn$_5$), T.T., H.P. (Bi$_2$O$_2$Se) and N.K., C.S., C.F. (Cd$_3$As$_2$). The experiment design, FIB microstructuring and the magnetotransport measurements were performed by C.G., C.P., A.E., K.R.S., M.B., P.J.W.M.. A.A. developed, applied and described the theoretical framework, and the analysis of experimental results has been done by C.G. and C.P.. Band structures were calculated by F.F., S.Z., Q.W., O.Y., C.F. and Y.S.. All authors were involved in writing the paper.\\
	\noindent \textbf{Competing Interests.} The authors declare no competing interests.\\

	\clearpage
	\begin {table}[t]
	\begin{center}
		\begin{tabular}{ | c | c | c | c | c |}
			\hline
			& \multicolumn{2}{|c|}{Topological} & \multicolumn{2}{|c|}{Schr\"odinger} \\ 
			
			\hline
			Materials  & $|E_F| = \big|\f{\partial E}{\partial (\log m_c)}\big|$   & $|E_F|  =2e\hbar F_0/m_c$ & $|E_F| = (1/2)|\f{\partial E}{\partial (\log g)}|$  & $|E_F|=e\hbar F_0/m_c$ \\ 
			\hline
			Cd$_3$As$_2$ & {278.9}   & 273 & 55.8 & 136.5 \\ 
			\hline
			LaRhIn$_5$ & 23.7 & 24 & / & 12  \\ 
			\hline
			Bi$_2$O$_2$Se & 201.9 & 42.8 & 20.2 & 21.4  \\ 
			\hline   
		\end{tabular}
		
		\caption {Self-consistency check for distinguishing topological or Schr\"odinger-type pocket. For each materials, the Fermi energy (in units of meV) is derived from $\Delta F /T^2$ and $F_0$ [\q{dmdE}] assuming the pocket is either Dirac- or Schr\"odinger-type {(see Supplementary Sec. II B). Here $g$ stands for the zero-field density of states.} The results clearly identify both Cd$_3$As$_2$ and LaRhIn$_5$ as topological materials, while for Bi$_2$O$_2$Se it clearly reveals its topologically-trivial nature.}
	\end{center}
\end{table}

\clearpage

\begin{figure}[t]
	\includegraphics[width=0.95\columnwidth]{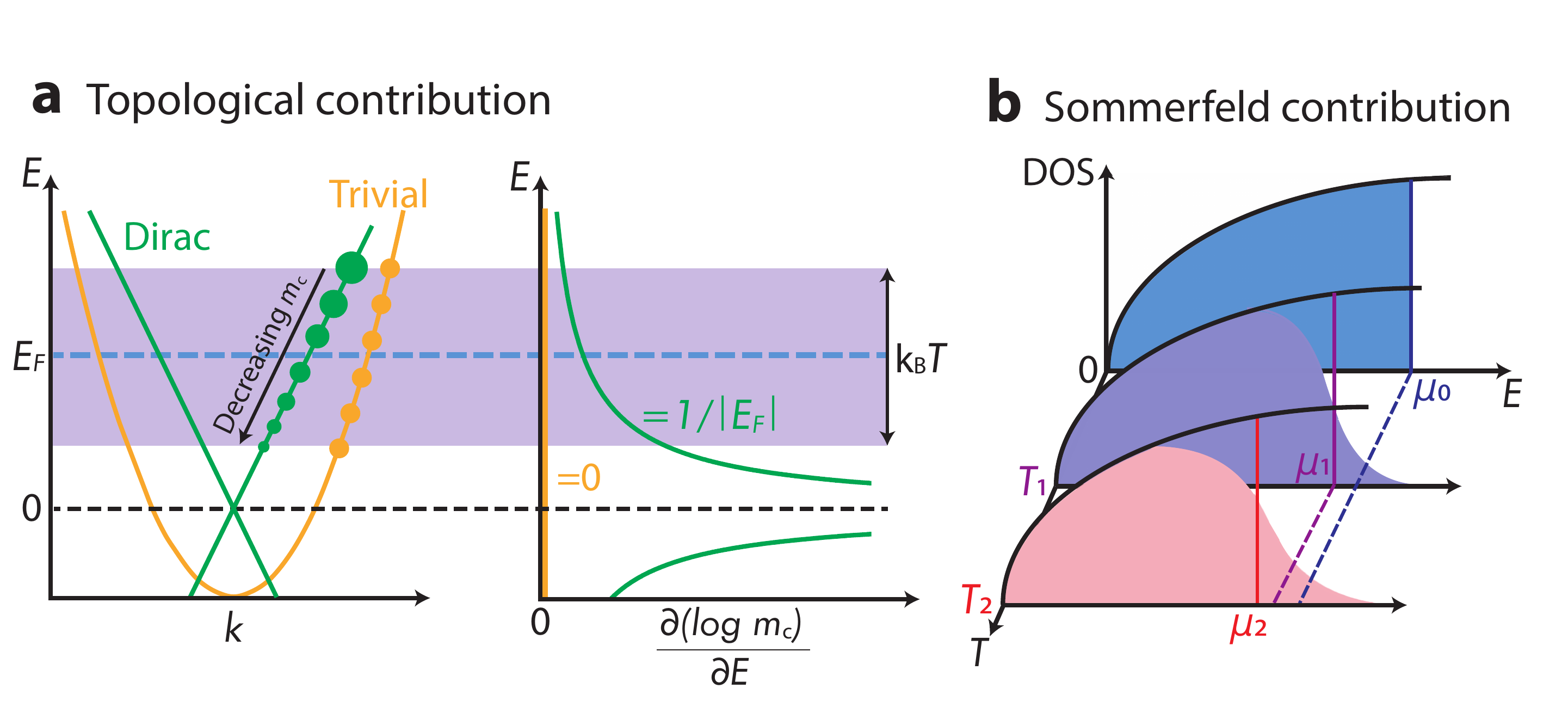}
	\caption{Illustration for topological and Sommerfeld contributions to temperature dependence of oscillation frequency. (a) For a linearly-dispersing Dirac-type pocket, the energy-derivative of the cyclotron mass, $\partial (\log m_c) / \partial E$ diverges when the Fermi level approaches the Dirac node. When approaching to the Dirac node,the Fermi pocket shrinks and the cyclotron mass is continuously decreasing to zero, therefore the smaller the oscillation frequency, the larger the oscillation amplitude. Due to the thermal broadening of chemical potential, this ultimately leads to the quadratic temperature dependence of the quantum-oscillation frequency. In constrast, for a  Schr\"odinger-type pocket with a parabolic dispersion, $\partial (\log m_c) / \partial E=0$. (b) Illustration of Sommerfeld contribution, it describes the shift of chemical potential at finite temperatures due to thermal broadening with a fixed carrier density.}
\end{figure}

\begin{figure}
	\includegraphics[width=0.45\columnwidth]{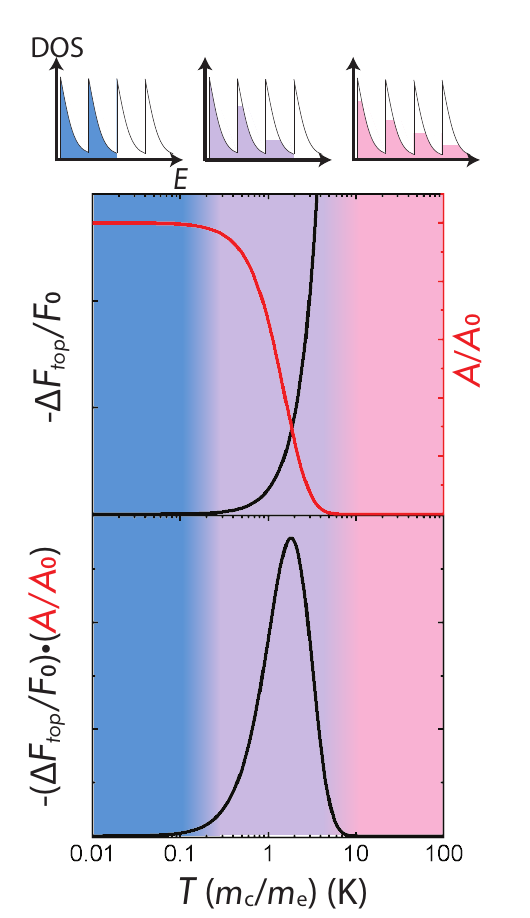}
	\caption{Optimal temperature range for detecting topological frequency shift.
		{The upper square panel displays the temperature dependence of the quantum oscillation amplitude ($\Delta A =A(T)-A_0$) and topological frequency shift $\Delta F_{top}(T)$. The temperature axis is scaled by the ratio of cyclotron to free-electron mass, and $F_0$ and $A_0$ stand for the frequency and amplitude at zero temperature, respectively. $-\Delta F_{top}/F_0$ steeply increases just before the oscillation amplitude vanishes; this corresponds to the temperature regime where thermal broadening is comparable to the cyclotron energy ($\var_c \approx 2\pi^2 k_BT$), as illustrated by the filled density-of-states (DOS) plot for various temperatures (at the very top of figure). The lower square panel plots the temperature dependence of $- (\Delta F_{top}/F_0) \cdot (A/A_0)$; its peak identifies the optimal temperature  to observe the topological frequency shift; our scaling of the temperature axis implies the optimal temperature is inversely proportional to the cyclotron mass.}}
\end{figure}

\begin{figure}
	\includegraphics[width=0.95\columnwidth]{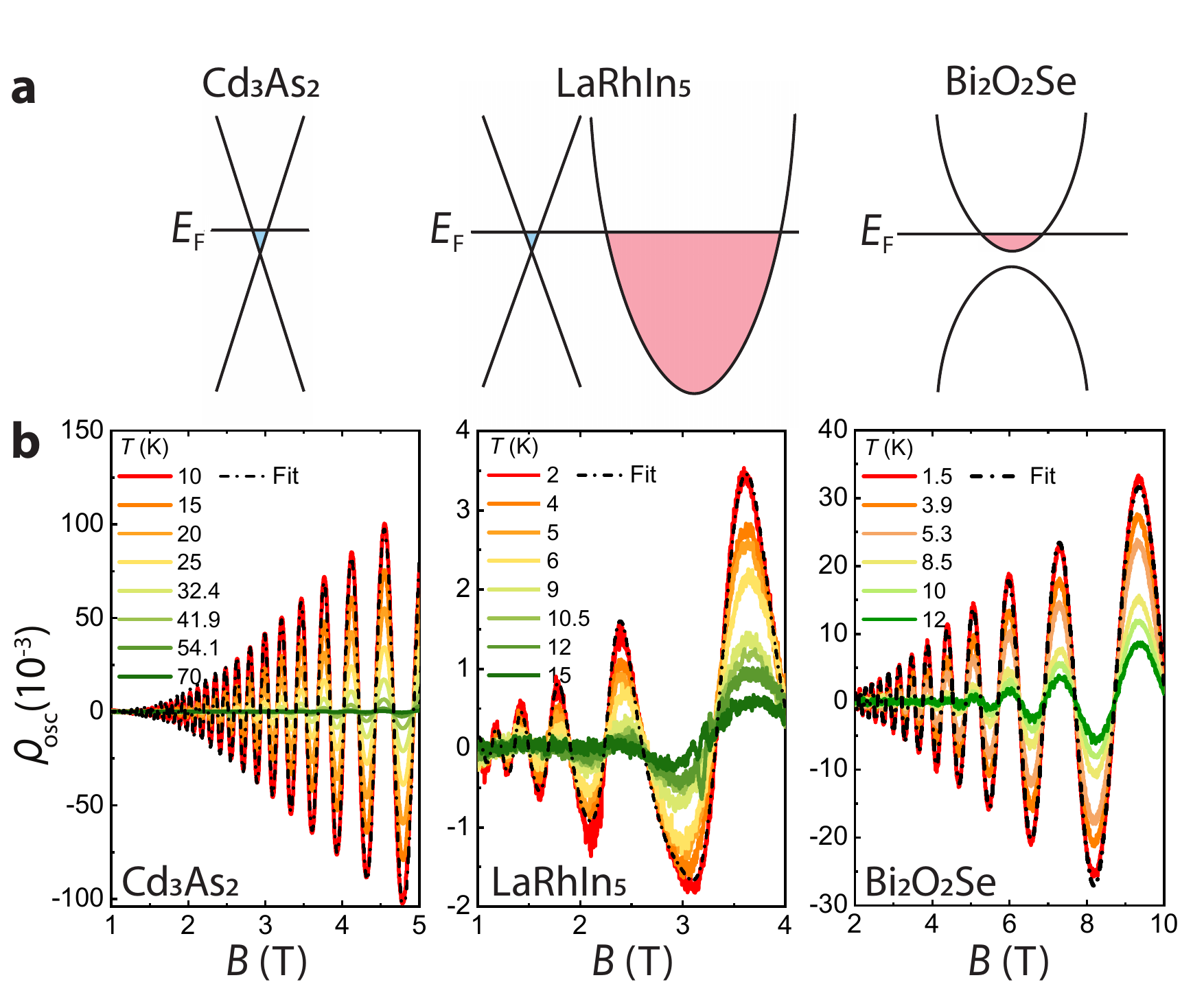}
	\caption{Experimental results of temperature-dependent quantum oscillation measurements. (a) Band structure illustration of three different types of materials, including Cd$_3$As$_2$ and Bi$_2$O$_2$Se where only one Fermi pocket and its symmetric copies sit at the Fermi level, as well as LaRhIn$_5$ where the small, candidate Dirac Fermi pocket coexists with large trivial pockets. (b) Temperature-dependent SdH oscillations of small Fermi pockets for Cd$_3$As$_2$ , LaRhIn$_5$ and Bi$_2$O$_2$Se respectively. {$\rho_{osc}=\Delta \rho / \rho_{BG}$, with $\Delta \rho$ the oscillatory part of the magnetoresistivity, and $\rho_{BG}$ a polynomial fit to the smooth background.} The dashed black line represents the Lifshitz-Kosevich fit to the quantum oscillation measured at lowest temperature for each material. {The magnetic field range is chosen to include as many low-noise quantum oscillations as possible, to improve the accuracy of frequency fitting. Exact device geometry and field/current orientations are described in the Supplementary Material.}}
\end{figure}

\begin{figure}
	\includegraphics[width=0.65\columnwidth]{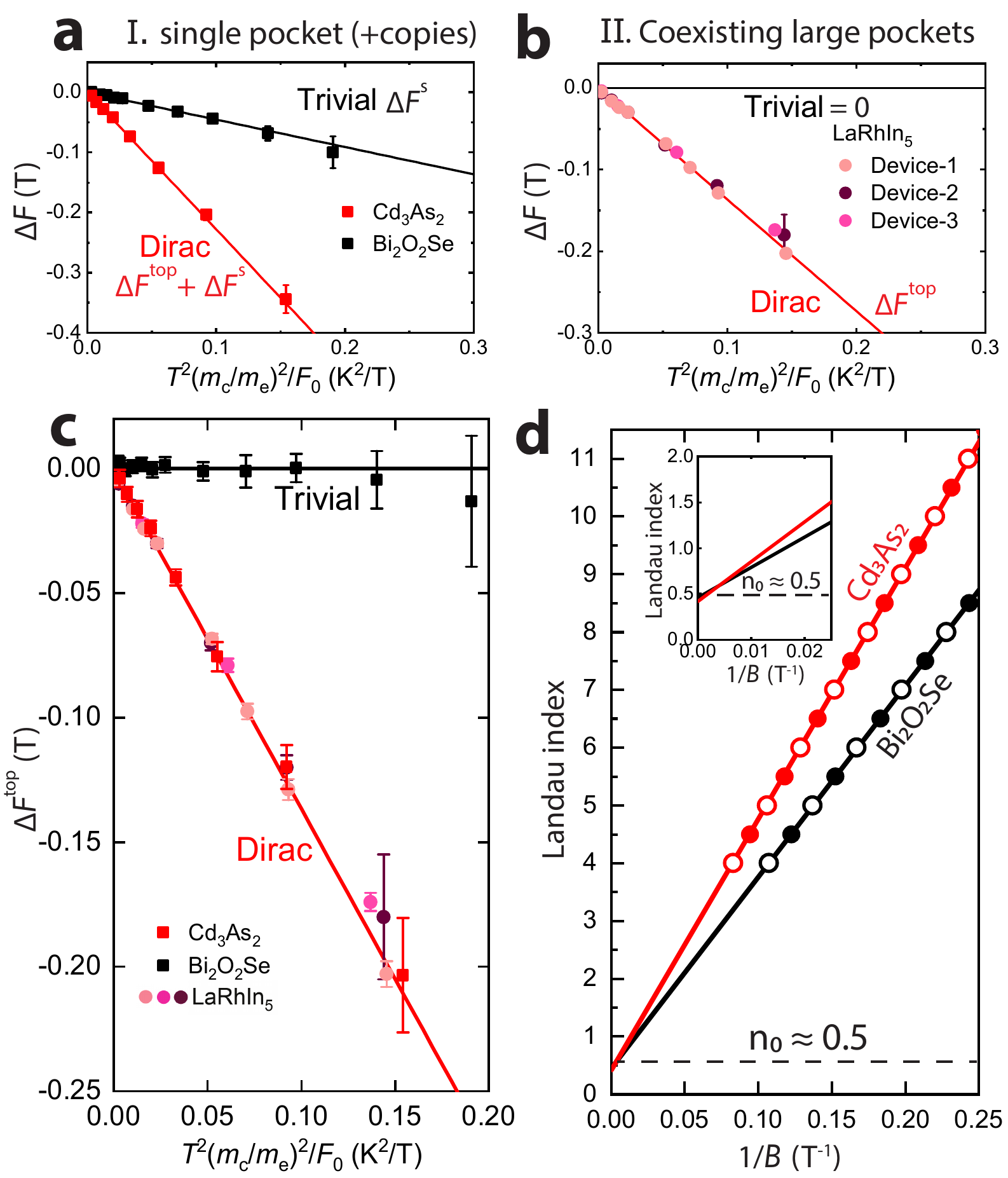}
	\caption{Analysis of frequency shift and Landau-fan plots. (a) presents the total frequency shift $\Delta F(T)$ versus  $T^2(m_c/m_e)/F_0$ for Cd$_3$As$_2$ and Bi$_2$O$_2$Se, both of which are semimetals with a single Fermi pocket plus symmetry-related copies (if any). (b) plots $\Delta F(T)$ for LaRhIn$_5$, which has a single, small pocket with coexisting large pockets. The solid lines display the fitting-parameter-free theoretical expectations $\Theta {(\pi k_BT)^2}/{\beta^2F_0}$. {For low carrier density semimetal, $\Theta = 5/48$ for Dirac case while for trivial case $\Theta = 1/48$. For Dirac metal with high carrier density $\Theta = 1/16$.} The filled squares and circles represent the experimentally determined value for different materials.  (c) displays the frequency shift after subtracting the Sommerfeld contribution ($\Delta F^s$); what remains is the topological frequency shift, if any. $\Delta F-\Delta F^s$ for both LaRhIn$_5$ (circles) and Cd$_3$As$_2$ (red squares) fall on the expected theoretical line (solid, red) for Dirac semimetal/metal, while for the topologically-trivial semimetal Bi$_2$O$_2$Se, the frequency shift nearly vanishes after subtraction. (d) Landau-fan plots of Cd$_3$As$_2$ and Bi$_2$O$_2$Se. The solid and empty symbols represents valleys and peaks of the quantum oscillations. Following the common interpretation of the residual Landau index (n$_0$ times $2\pi$) as a quantized Berry phase, one would erroneously diagnose both Cd$_3$As$_2$ and Bi$_2$O$_2$Se as topologically non-trivial. All error bars in this figure are determined by the standard error of the fitting parameters generated by the non-linear regression fitting procedure.}
\end{figure}

\end{document}


\renewcommand{\theequation}{\arabic{equation}}
	\newcommand{\beginsupplement}{%
		\setcounter{table}{0}
		\renewcommand{\thetable}{\arabic{table}}%
		\setcounter{figure}{0}
		\renewcommand{\thefigure}{\arabic{figure}}%
	}
	\newcommand{\lk}{Lifshitz-Kosevich }
	\newcommand{\equref}[1]{Eq. (\ref{#1})}
	
	\title{Supplemental Material for\\``Temperature dependence of quantum oscillations from non-parabolic dispersions"}
	
	\author{Chunyu Guo$^{\ast\dagger}$}\affiliation{Laboratory of Quantum Materials (QMAT), Institute of Materials (IMX), \'{E}cole Polytechnique F\'{e}d\'{e}rale de Lausanne (EPFL), CH-1015 Lausanne, Switzerland}
	\author{A. Alexandradinata$^{\ast\dagger}$}\affiliation{Institute for Condensed Matter Theory, University of Illinois at Urbana-Champaign, Urbana, Illinois 61801, USA}\affiliation{Department of Physics, University of Illinois at Urbana-Champaign, Urbana, Illinois 61801, USA}
	\author{Carsten Putzke}\affiliation{Laboratory of Quantum Materials (QMAT), Institute of Materials (IMX), \'{E}cole Polytechnique F\'{e}d\'{e}rale de Lausanne (EPFL), CH-1015 Lausanne, Switzerland}
	\author{Amelia Estry}\affiliation{Laboratory of Quantum Materials (QMAT), Institute of Materials (IMX), \'{E}cole Polytechnique F\'{e}d\'{e}rale de Lausanne (EPFL), CH-1015 Lausanne, Switzerland}
	\author{Teng Tu}\affiliation{Center for Nanochemistry, Beijing National Laboratory for Molecular Sciences (BNLMS),College of Chemistry and Molecular Engineering, Peking University, Beijing 100871, China.}
	\author{Nitesh Kumar}\affiliation{Max Planck Institute for Chemical Physics of Solids, 01187 Dresden, Germany}
	\author{Feng-Ren Fan}\affiliation{Max Planck Institute for Chemical Physics of Solids, 01187 Dresden, Germany}
	\author{Shengnan Zhang}\affiliation{Chair of Computational Condensed Matter Physics (C3MP), Institute of Physics (IPHYS), \'{E}cole Polytechnique F\'{e}d\'{e}rale de Lausanne (EPFL), CH-1015 Lausanne, Switzerland}\affiliation{National Centre for Computational Design and Discovery of Novel Materials MARVEL,\'{E}cole Polytechnique F\'{e}d\'{e}rale de Lausanne (EPFL), CH-1015 Lausanne, Switzerland}
	\author{Quansheng Wu}\affiliation{Chair of Computational Condensed Matter Physics (C3MP), Institute of Physics (IPHYS), \'{E}cole Polytechnique F\'{e}d\'{e}rale de Lausanne (EPFL), CH-1015 Lausanne, Switzerland}\affiliation{National Centre for Computational Design and Discovery of Novel Materials MARVEL,\'{E}cole Polytechnique F\'{e}d\'{e}rale de Lausanne (EPFL), CH-1015 Lausanne, Switzerland}
	\author{Oleg V. Yazyev}\affiliation{Chair of Computational Condensed Matter Physics (C3MP), Institute of Physics (IPHYS), \'{E}cole Polytechnique F\'{e}d\'{e}rale de Lausanne (EPFL), CH-1015 Lausanne, Switzerland}\affiliation{National Centre for Computational Design and Discovery of Novel Materials MARVEL,\'{E}cole Polytechnique F\'{e}d\'{e}rale de Lausanne (EPFL), CH-1015 Lausanne, Switzerland}
	\author{Kent R. Shirer}\affiliation{Max Planck Institute for Chemical Physics of Solids, 01187 Dresden, Germany}
	\author{Maja D. Bachmann}\affiliation{Max Planck Institute for Chemical Physics of Solids, 01187 Dresden, Germany}\affiliation{School of Physics and Astronomy, University of St Andrews, St Andrews KY16 9SS, UK}
	\author{Hailin Peng}\affiliation{Center for Nanochemistry, Beijing National Laboratory for Molecular Sciences (BNLMS),College of Chemistry and Molecular Engineering, Peking University, Beijing 100871, China.}
	\author{Eric D. Bauer}\affiliation{Los Alamos National Laboratory, Los Alamos, New Mexico 87545, USA}
	\author{Filip Ronning}\affiliation{Los Alamos National Laboratory, Los Alamos, New Mexico 87545, USA}
	
	\author{Yan Sun}\affiliation{Max Planck Institute for Chemical Physics of Solids, 01187 Dresden, Germany}
	\author{Chandra Shekhar}\affiliation{Max Planck Institute for Chemical Physics of Solids, 01187 Dresden, Germany}
	\author{Claudia Felser}\affiliation{Max Planck Institute for Chemical Physics of Solids, 01187 Dresden, Germany}
	\author{Philip J. W. Moll$^{\dagger}$}\affiliation{Laboratory of Quantum Materials (QMAT), Institute of Materials (IMX), \'{E}cole Polytechnique F\'{e}d\'{e}rale de Lausanne (EPFL), CH-1015 Lausanne, Switzerland}

	\date{\today}
	\maketitle
	
	\beginsupplement
	
	{\tableofcontents \par}
	\clearpage
	
	\section*{Supplementary Note 1: Review of quantization rule and standard Lifshitz-Kosevich theory}
	
	\subsection*{A. Review of Onsager-Lifshitz-Roth quantization rule}
	
	The phase that an electron accumulates as it makes a cyclotron orbit around the Fermi surface must be an integer multiple of $2\pi$, according to the Onsager-Lifshitz-Roth quantization rule\cite{onsager,lifschitz1954theory,Roth_SOCandZeeman,phiB_Aris_PRX,100page}:
	
	\e{
		l_B^2 S + \pi+\lambda_v^s  = 2\pi n , \;\;n \in \mathbb{Z},\la{qrule}
	}
	with $l_B = (h /e|B|)^{1/2}$  the magnetic length and $S$ describes a cross-sectional area of the Fermi pocket. The subdominant term includes a quantum Maslov correction ($\pi$ for orbits deformable to a circle \cite{100page}), and a \textit{quantization offset}  $\lambda_v^s$ that encodes the quantum geometry of wave functions on the Fermi surface (the Berry phase), and also the generalized Zeeman interaction of the intrinsic spin and orbital magnetic moment\cite{rothI,Roth_SOCandZeeman,NQOrbital,Zeeman+SOCphi_GeoPhi_PNAS}. 
	
	We have introduced an index $s=\uparrow,\downarrow$ which distinguishes orbits which are spin-degenerate; $s$ corresponds to intrinsic spin in the absence of spin-orbit coupling, but otherwise should be interpreted as a pseudospin.  For spin-split Fermi surfaces, any individual orbit with a given cross-sectional area is non-degenerate, hence the index $s$ is unnecessary. 
	
	The valley index $v=1..N_{val}$ distinguishes between symmetry-related orbits that are separated in $k$-space.
	In this work, we focus on time-reversal-symmetric, 3D metals which may have a crystalline center of inversion; both time-reversal and spatial-inversion symmetries invert $\bk\rightarrow -\bk$, thus if an orbit is not centered around an time-reversal-invariant momentum (TRIM), then it necessarily has a time-reversed-partner orbit in a different valley that encloses the same area $S$, as illustrated in Fig. 1 of the main text. The same figure also summarizes the total number of orbits (counting both spin and valley degrees of freedom) in each symmetry class of orbits that we consider.
	
	\subsection*{B. Lifshitz-Kosevich formula for de Haas-van Alphen oscillations}
	
	For 3D metals, the Fermi energy may be considered fixed  up to small corrections $\propto  B^{3/2}$\cite{shoenberg2009magnetic}. 
	The  Onsager-Lifshitz-Roth quantization rule [cf.\ Supplementary \equref{qrule}] gives the field  $B_n$ at which the $n$'th Landau level crosses a fixed Fermi level;  such crossings occur periodically in $1/B$ with frequency $F_0(E_F)=\hbar  S(E_F)/2\pi e$, leading to  oscillations in  thermodynamic state variables such as magnetization, known as the de Haas-van Alphen effect\cite{dHvA}. For 3D non-magnetic metals, the oscillatory component of magnetization has the form
	\cite{Roth_SOCandZeeman,phiB_Aris_PRX}:
	\begin{equation}
	\Delta M \propto \sqrt{B}\sum_{r=1}^{\infty}\cos(r\lambda) \;\f{R_D^r R_T^r}{r^{3/2}} \sin\left[ r\left(2\pi \f{F_0}{B}+\pi\right) +\phi_{LK}\right]. \la{oscillatorymag}
	\end{equation}
	For brevity, we have omitted a proportionality constant that is field- and temperature-independent; this constant is irrelevant for the purpose of topofermiology. 
	Above, $\phi_{LK}=+\pi/4$ ($-\pi/4$) for a minimal and  maximal orbit respectively\cite{lifshitz_kosevich_jetp}. While Supplementary \equref{oscillatorymag} sums oscillations from all symmetry-related orbits (counting spin and valley degrees of freedom) with possibly distinct offsets $\lambda_s^v$, the symmetry analysis of \ocite{phiB_Aris_PRX} shows that the time reversal symmetry imposes a zero-sum rule: $\sum_{s,v}\lambda_s^v=0$ mod $2\pi$, and moreover there is only one independent $\lambda_s^v$ which we will denote by $\lambda$; $\lambda$ enters Supplementary \equref{oscillatorymag} as a cosine pre-factor. The presence of higher harmonics ($r>1$) results from the Landau-level density of states being sharply peaked. This sharpness is partially smeared due to the finite quasiparticle lifetime  $\tau$ and nonzero temperature $T$, which result in the well known Dingle amplitude factor and thermal damping factor respectively:
	\begin{equation}
	R_D^r:=\exp\bigg({-}\f{r}{2}\f{h}{\var_c\tau}\bigg),\as R_T^r:=\f{a_r}{\text{sinh}\,(a_r)},\as a_r:= \f{2\pi^2 r k_BT}{\var_c}, \la{thermalfactor}
	\end{equation}
	with $\var_c=\hbar |eB|/m_c $ the cyclotron energy,  $m_c=(2\pi)^{-1}\partial S/\partial E$ the cyclotron mass, and $\tau$ the quantum lifetime.

	\subsection*{C. Lifshitz-Kosevich formula for Shubnikov-de Haas oscillations}

	Compared to thermodynamic oscillations, quantum oscillations in transport coefficients, known as the Shubnikov-de Haas effect\cite{SdH},  are typically harder to interpret owing to multiple scattering mechanisms. We focus only on the subset of transport coefficients whose quantum oscillations plausibly have the Lifshitz-Kosevich form, so that standard routines exist for extracting Fermi-surface cross sections and cyclotron masses in the Lifshitz-Kosevich phenomenology.
	
	In this regard, the longitudinal magnetoconductance is more amenable to Lifshitz-Kosevich analysis than the transverse magnetoconductance. Oscillations of the latter coefficient are known to have non-Lifshitz-Kosevich corrections due to intra-Landau-level scattering -- this becomes significant for lower Landau levels and for higher harmonics\cite{RothArgyres}.
	
	Though the longitudinal magnetoconductance is not directly measured, its value may be inferred from a measurement of the longitudinal magnetoresistance -- assuming the field-dependent resistivity tensor elements $\rho_{i\parallel B,j \perp B}$ are small compared to other elements; here, $i$ lies in the field direction but $j$ does not, and these elements are sometimes called the planar Hall elements. (This assumption certainly holds for isotropic bands\cite{Galvanomagnetic_Effects}, and may be checked empirically on a case-by-case basis, as exemplified for LaRhIn$_5$ in \s{sec:longitudinalres}.)  Then the longitudinal resistivity reduces to the inverse of the longitudinal conductivity: $\sigma_{\parallel}\approx 1/\rho_{\parallel}.$ The reason is that the conductivity tensor becomes block-diagonal with respect to longitudinal and transverse directions, and the longitudinal block is one-by-one and trivially inverted. It follows that the ratio of  quantum-oscillatory to semiclassical (non-oscillatory) components satisfy: 
	\begin{equation}
	\f{\Delta \rho_{\parallel}}{\rho^0_{\parallel}}= - \f{\Delta \sigma_{\parallel}}{\sigma_{\parallel}^0} +O\bigg(\bigg[\f{\Delta \sigma_{\parallel}}{\sigma_{\parallel}^0}\bigg]^2\bigg).\la{ratios}
	\end{equation}
	For the longitudinal magnetoconductance, it is known for certain classes of Fermi surfaces that the effect of the magnetic field on localized-impurity scattering is given entirely through its effect on the density of states, whose oscillatory behavior is known to have a Lifshitz-Kosevich form. Consequently, $\Delta \rho_{\parallel}/\rho^0_{\parallel}$ would also have a Lifshitz-Kosevich form:
	\begin{equation}
	\Delta\rho_{\parallel}/\rho^0_{\parallel} \propto \sqrt{B}  \sum_{r=1}^{\infty}  \;\f{R_D^r R_T^r}{r^{1/2}}\cos(r\lambda)\, {\cos\left[ r\left(2\pi \f{F_0}{B}+\pi\right) +\phi_{LK}\right]}.
	\la{sdhformula}
	\end{equation}
	This formula has been derived for parabolic bands in the effective-mass approximation, and is also valid for scattering off acoustic phonons at high temperatures (compared to the phonon energy)\cite{argyres1958quantum,RothArgyres}. To our present knowledge, the explicit derivation of Supplementary \equref{sdhformula} for Dirac-Weyl Fermi surfaces does not yet exist and would represent a significance advance.

	\section*{Supplementary Note 2: Extended Lifshitz-Kosevich theory for high-temperature quantum oscillations}

	\subsection*{A. Frequency correction for high-temperature quantum oscillations}
	
	The Lifshitz-Kosevich formulae in Supplementary \equref{oscillatorymag} and Supplementary \equref{sdhformula} are leading terms of an asymptotic expansion in powers of $k_BT/E_F$ -- the degeneracy parameter of Fermi gases. We now extend the Lifshitz-Kosevich formula to the next order in $k_BT/E_F$, to manifest the energy dependence of the cyclotron mass: $\partial m_c/\partial E$. 
	
	Following a natural generalization of Roth's derivation in \ocite{Roth_SOCandZeeman}, this extension amounts to replacing $R_T^r   \sin[r(2\pi F_0/B+\pi )\pm \phi_{LK}]$ in Supplementary \equref{oscillatorymag} by 
	\begin{equation}
	\text{Imag}\left\{  e^{ir(2\pi F_0/B +\pi )+i\phi_{LK}}\sum_{p=0}^{\infty} 2a_r e^{-a_r(2p+1)-ib_r(2p+1)^2} \right\};\as b_r:=\pi^3 r\f{(kT)^2}{\var_cm_c}\bigg|\p{m_c}{\var}\bigg|,\la{replace}
	\end{equation}
	and $a_r$ defined in Supplementary \equref{thermalfactor}; note $b_r/a_r \sim k_BT/E_F$ for the Dirac-Weyl fermion.
	
	While this form holds for arbitrary ratios of $k_BT$ vs the cyclotron energy (with the constraint $k_BT/E_F$ is small), it is typically the high-temperature regime (defined by $2\pi^2 k_BT/\var_c \gtrsim 1$) that is  most amenable to experimental analysis. In this regime where the high Landau levels are smeared out by temperature, only the fundamental harmonic ($r=1$) and the $p=0$ term in Supplementary \equref{replace} survives, hence  the sole correction to the \lk formula is a $T^2$-correction to the oscillation frequency:
	\begin{equation}
	F_0(\mu)\rightarrow {F}(\mu,T)=F_0(\mu)- \f{\pi^2}{4}\f{(k_BT)^2}{\beta} \bigg|\f{\partial (\log m_c)}{\partial E}\bigg|,\la{correctF}
	\end{equation}
	with $\mu$ the chemical potential and $\beta:=e\hbar/2m_cc$ the effective Bohr magneton. The above effective-mass correction to the frequency holds generally for any Fermi surface; when applied to the Dirac-type Fermi surface, the result matches a calculation by K\"uppersbusch and Fritz for 2D Dirac systems\cite{KupperbuschFritz}; however their proposed \lk formula for 2D Dirac systems is discontinuous in $T$ and $\mu$, owing to a choice of integration path (vertical line in Fig. 2 of \ocite{KupperbuschFritz}) that non-generically intersects poles of their integrand. In comparison, the \lk formula that is presented here is continuous in $T$ and $\mu$. An analogous frequency correction was also calculated for graphene by Fortin and Audouard\cite{Fortin_Graphene}, however their proportionality factor for the $T^2$-correction is erroneously larger by a factor of two.
	
	Beside the explicit $T$-dependence of $F(\mu,T)$  through the correction term in Supplementary \equref{correctF}, there is additionally an implicit $T$-dependence through $\mu(T)$ for systems of fixed particle density. The leading term of a Sommerfeld expansion gives
	\begin{equation}
	\mu(T)= E_F- \frac{1}{6} \,(\pi k_BT)^2\,\frac{g'(E_F)}{g(E_F)}+O(T^4),\la{sommerfeld}
	\end{equation}
	where $g(\var)$ is the zero-field density of states, with $g'$ the derivative of $g$ with respect to energy. The effects of Supplementary \qq{correctF}{sommerfeld} are combined for a topo-fermiological analysis in \s{sec:consistency}.
	
	Other possible mechanisms of temperature dependence are reviewed in \ocite{shoenberg2009magnetic}.  Electron-phonon interactions and thermal expansion of the solid (as contributed by lattice vibrations) are known to lead to a $T^4$-dependence\cite{lonzarich1983temperature}, which is separable (in data analysis) from the $T^2$-dependence identified in Supplementary \qq{correctF}{sommerfeld}. Barring ferromagnetic metals\cite{lonzarich1974temperature}, we know of only one other mechanism which in principle leads to a $T^2$-dependence: namely, the entropic contribution of Fermi-liquid quasiparticles gives a linear-in-$T$ term in the thermal expansion tensor\cite{barron1980thermal}, which implies a $T^2$ correction to $F$ -- assuming that the change in $F$ is linearly related to the change in volume of solid (or its strain parameter). In most metals, the thermal expansion is dominantly driven by lattice, the oscillation frequency therefore shows a clear $T^4$ temperature dependence, and the changes in frequency are at most parts in $10^4$ over the temperature range that quantum oscillations are observable\cite{shoenberg2009magnetic}. There exists one known case of a significantly larger effect for the `needle' pocket of Zinc\cite{PhysRev.95.1421} where both electronic and lattice contributions are observable, evident by the temperature dependence of oscillation frequency which contains both the $T^2$ and $T^4$ terms. This rare example demonstrates that for nearly-free-electron metals, the dimensions of small pockets near the Brillouin-zone boundary can be disproportionately sensitive to small changes in lattice dimensions\cite{PhysRev.151.484}. Therefore when  oscillation frequency no longer shows a simple $T^2$-dependence, one needs to be careful of attributing the frequency change to a topological origin.

	
	\subsection*{B. Consistency relations for  topo-fermiology}\la{sec:consistency}
	
	Assuming that the effect of thermal expansion is negligible, we derive consistency relations between the $T^2$ coefficient of $F$ and its zero-temperature extrapolation -- these relations being different for parabolic and linear bands allow to distinguish between Schr\"odinger- and Dirac-type pockets. 
	
	Let us first consider the case where a small Fermi pocket under study coexists with other Brillouin-zone-sized pockets. Then an order of magnitude estimate for $g'/g$ in  Supplementary \equref{sommerfeld} is given by the inverse of the bandwidth $E_{bw}$, which is typically large compared to both $k_BT$ (the standard degeneracy condition) and the Fermi energy $E_F$ of the small pocket (measured from band bottom or the Dirac-Weyl node, depending on context). This implies the Sommerfeld correction of $\mu$ is small compared to the Fermi energy:
	$\delta \mu/E_F \sim (k_BT)^2/(E_F E_{bw})$. The Sommerfeld correction  to the frequency is  of the order of  $(k_BT)^2/(\beta E_{bw})$, with $\beta$ the effective Bohr magneton of the small pocket; in comparison, the frequency correction to the frequency [Supplementary \equref{correctF}] is larger by a factor of $E_{bw}/E_F$ for the Dirac-type pocket {with linear dispersion.} For these reasons, we ignore the Sommerfeld correction to $\mu$ in our analysis of the multi-band metal LaRhIn$_5$. Substituting {Eq.\ (1)} of the main text into Supplementary \equref{correctF} and replacing $E_F$ with $|E_F|  =2e\hbar F_0/m_c$ for a Dirac pocket, we derive for the Dirac-type pocket a consistency relation between the $T^2$ coefficient of $F$ and its zero-temperature extrapolation:
	\begin{equation}
	\text{{Linear; coexisting large pocket:}}\as {F}(\mu,T)-F_0\approx \Delta F^{top}= - \f{1}{16}\f{(\pi k_BT)^2}{\beta^2 F_0}.\la{consistencymufixed}
	\end{equation}
	
	Lastly, we consider semimetals with either a single, small Fermi pocket encircling a TRIM or a pair of time-reversal-related pockets as applicable to Cd$_3$As$_2$ and Bi$_2$O$_2$Se. Assuming there are no other Fermi surfaces, the oscillation contains only a single frequency and its higher harmonics. Then the Sommerfeld correction to frequency is comparable to the effective-mass correction, and should be accounted for in deriving a different consistency relation. In the case of the parabolic band, $g'/g=(d/2-1)/E_F$ in $d$ spatial dimensions, leading to a nontrivial frequency correction for $d=3$:
	\begin{equation}
	\text{{Parabolic; single frequency:}} \as F(\mu,T) -F_0= \Delta F^s= - \frac{1}{48} \f{(\pi k_BT)^2}{\beta^2F_0 }+O(T^4).\la{Fparabolic}
	\end{equation}
	In the case of linear bands, $g'/g=(d-1)/E_F$, therefore the Sommerfeld correction can be written as:
	\begin{equation}
	\text{{Linear; Sommerfeld correction:}}\as \Delta F^s = - \frac{1}{24} \f{(\pi k_BT)^2}{\beta^2F_0 }+O(T^4);\la{DiracSommerfeld}
	\end{equation}
	for $d=3$. Summing the Sommerfeld correction with the effective-mass correction, the net frequency shift for linear bands is
	\begin{equation}
	\text{{Linear; single frequency:}}\as F(\mu,T) -F_0= - \frac{5}{48} \f{(\pi k_BT)^2}{\beta^2F_0 }+O(T^4).\la{consistencymuvary}
	\end{equation}
	We see that the coefficient of the $T^2$ term is larger by a factor of five compared to the parabolic case [cf.\ Supplementary \equref{Fparabolic}].
	
	\subsection*{C. Applicability to heavy-fermion materials} \la{sec:mstardiscussion}
	
	Our method is generally applicable to materials with large cyclotron mass. First of all, 
	$\Delta F/F_0$ gives relative shift of frequency. For this shift to be observable, amplitude of oscillations must not be overly diminished by thermal damping. Hence we consider the product $-(\Delta F/F_0)(A/A_0)$. 
	As shown in Fig. 3, it becomes significant at elevated temperatures just before quantum oscillations vanish. This demonstrates that heavy quasiparticle mass does not preclude the observation of oscillation frequency change, it simply renormalizes the temperature scale to lower values just as it does for the quantum oscillations themselves.
	
	The physical origin for the general applicability is that both the thermal damping of the quantum oscillation amplitude, $A(T)$, and the strength of the frequency shift, $\Delta F(T)^{top}$, are set by the same physics. Key is to compare the Landau level spacing, i.e. the cyclotron energy, to the thermal broadening of the Fermi-Dirac distribution. Therefore the cyclotron mass value only sets the temperature range at which the oscillation frequency change is best resolvable (Supplementary \fig{fig:mass}). Importantly, if quantum oscillations in a heavy fermion material can be observed at sufficient amplitude, $\Delta F(T)^{top}$ can be observed as well.
	
	\subsection*{D. Distinguishing between 3D Dirac points and  nodal lines} \la{sec:distinguishpointline}
	
	Given that a pocket is of Dirac type, one may ask a more fine-grained question of the Dirac crossing: does the cyclotron orbit (a) encircle an isolated $\bk$-point (3D Dirac point) of four-fold energy degeneracy, or does it (b) link with an isolated $\bk$-point (nodal line) of four-fold energy degeneracy?
	In the former case, the energy degeneracy splits linearly in $\bk$ along all directions away from the Dirac point. In case (b), the energy degeneracy only splits linearly in two directions that are locally orthogonal to the nodal line; there is therefore a 2D Dirac fermion in any small $\bk$-rectangle pierced by the line. 
	
	To distinguish between these scenarios, it is worth reviewing the role of crystallographic symmetry and the spin-orbit interaction in stabilizing either type of degeneracies. In non-magnetic, centrosymmetric metals with negligible spin-orbit coupling, it is known that band degeneracies occur along lines, and such degeneracies are generically lifted by the spin-orbit interaction. 
	According to degenerate perturbation theory, the energy splitting occurs to the first order in the spin-orbit coupling; in contrast, spin-orbit-induced shifts of  energy levels (away from the nodal line) occur to the second order.

	Suppose a cyclotron orbit is found that satisfies either Supplementary \equref{consistencymufixed} or Supplementary \equref{consistencymuvary} depending on context,  implying that the pocket is linearly dispersing in the two directions orthogonal to the field.  If we further hypothesize that this orbit links with a nodal line, we must presume that the spin-orbit splitting $E_{soc}$ of the nodal line is so weak as to not spoil this linear dispersion. A necessary consistency condition is that
	\begin{equation}
	E_{soc}\ll |E_F|=\bigg|\f{\partial E}{\partial \log m_c}\bigg|,\la{socconsistency}
	\end{equation}
	with $|E_F|$ measured from the four-fold degeneracy (within the small $\bk$-rectangle pierced by the line); the second equality in Supplementary \equref{socconsistency} follows from the linearity of the Dirac dispersion, and note $\partial E/\partial \log m_c$ is directly measurable from the $T^2$ dependence of the oscillation frequency.
	
	On the other hand, certain crystallographic point-group symmetries (beyond spatial inversion) stabilize four-fold degeneracies at isolated points in $\bk$-space, no matter the strength of the  spin-orbit interaction. For concreteness, we focus on rotational symmetry of order greater than two, which is a symmetry of the 3D Dirac semimetals Cd$_3$As$_2$\cite{zhijun_cd3as2} and Na$_3$Bi\cite{zhijun_na3bi}, as well as the candidate LaRhIn$_5$.
	Assuming such symmetry,  a Dirac-type cyclotron orbit  may either encircle a Dirac point or link with a nodal line (cases a,b). If the spin-orbit energy scale is found to be comparable or greater than the experimentally attained $|E_F|$, then the nodal-line scenario  is ruled out according to 
	Supplementary \equref{socconsistency}, and the Dirac-point scenario remains as the only consistent scenario. 
	
	Unfortunately we know of no direct means to estimate $E_{soc}$ from quantum oscillations; such quantity can in principle be measured by angle-resolved photoemission spectroscopy, or estimated from first principles -- it is now routine to compare band structures (of the same material) with and without spin-orbit coupling. 
	Such a first-principles determination of $E_{soc}$ will be used to argue for the Dirac-point scenario for LaRhIn$_5$, in \s{sec:pointvslines_larhin} below.
	
	\section*{Supplementary Note 3: Analysis of temperature-dependent oscillation frequency for the experimental case studies}
	
	\subsection*{A. Lifshitz-Kosevich analysis of the longitudinal magnetoresistance oscillations}\la{sec:longitudinalres}\la{sec:fit}

	Supplementary \fig{fig:Rtensor} shows that for LaRhIn$_5$ the planar-Hall elements (of the longitudinal resistivity) are significantly smaller than all other elements, thus justifying Supplementary \equref{ratios} and the Lifshitz-Kosevich form of the longitudinal magnetoconductance,
	
	\begin{equation}
	\begin{aligned}
	&\f{\Delta \rho_{\parallel}}{\rho^0_{\parallel}} = \sum_{r = 1}^{\infty} A' \sqrt{\frac{B}{r}}  cos(r\lambda_1^{\uparrow}) \cdot  \frac{\alpha rm_cT/B}{sinh(\alpha rm_cT/B)} \cdot e^{-\frac{1140r\sqrt{F}}{l_{q}B}} \cdot cos[r(2\pi \frac{F}{B}-\pi)+\phi_{LK}]
	\la{truefit}
	\end{aligned}
	\end{equation}
	
	Here $\alpha = 2\pi^2ck_B/e\hbar = 14.69~T/K$, $r$ is the order of harmonics, $F$ corresponds to the oscillation frequency, $l_q$ represents the quantum mean free path of electrons in cyclotron motion, $m_c$ stands for the cyclotron mass in unit of free electron mass, and $\phi_{LK}$ = $\pi/4$ is the correction phase for a minimal orbit of a 3D Fermi surface. Note that for the Dingle damping term ($e^{-\frac{1140r\sqrt{F}}{l_{q}B}}$) we assume the Fermi surface is isotropic since the oscillation frequency of both Cd$_3$As$_2$ (Supplementary \fig{fig:anglepolar-2}) and LaRhIn$_5$ is almost angle-independent (Supplementary \fig{fig:anglepolar}). For Bi$_2$O$_2$Se, the Fermi surface is ellipsoidal as evident by the clear change of oscillation frequency with field applied along different crystal axis (Supplementary \fig{fig:anglepolar-2}). In this case, the magnetic field is applied along the ellipsoidal axis ($B\parallel c$) to avoid possible complexity due to Fermi surface anisotropy.
	The entire experimental dataset measured at different temperatures is then fitted globally by a standard least squares regression method using the non-linear model fitting function provided by Mathematica. In this procedure, both the cyclotron mass ($m_c$) and the quantum mean free path ($l_q$) are set to be temperature-independent while the oscillation frequency ($F$) is $not$ restricted to the same value at different temperatures, which directly yields the best-fit $\Delta F(T)$ to all experimental data. The error bar is determined by the standard error of the fitting parameters generated by the non-linear regression fitting procedure. The value and error bar of the fitting parameters for all three materials are listed in Table (S1). Note that for Cd$_3$As$_2$ and Bi$_2$O$_2$Se the quantization offset value cannot be determined due to the absence of higher harmonic oscillations.
	
	\par
	\begin {table}[H]
	\begin{center}
		\begin{tabular}{ | c | c | c | c | c |}
			\hline
			Materials  & Cd$_3$As$_2$  & LaRhIn$_5$ (Device-1) & Bi$_2$O$_2$Se \\ \hline
			$A'$~($10^{-3}$)  & 118.1$\pm$4.6 & 2.5$\pm$0.06  & 36$\pm$0.03 \\ \hline
			$\lambda_1^\uparrow~(\pi)$ & / & 0.98$\pm$0.01 & /  \\ \hline
			$F_0$~(T) & 43.75$\pm$0.004 & 6.94$\pm$0.001 & 33.28$\pm$0.003 \\ \hline  
			$\frac{dF}{d(T^2)}$~(10$^{-4}$ T/K$^2$) & -0.701$\pm$0.003 & -8.94$\pm$0.12 & -5.02$\pm$0.1 \\ \hline  
			$m_c~(m_e)$ & 0.037$\pm$0.003 & 0.067$\pm$0.007 &  0.18$\pm$0.008 \\ \hline
			$l_q$~(nm)  & 230.7$\pm$23 & 129$\pm$7 & 130.7$\pm$20\\ \hline   
		\end{tabular}
		
		\caption {List of values and standard errors for all fitting parameters for different materials. {Here parameter d$F$/d$T^2$ (A) is produced by the polynomial fitting of temperature-dependent $F(T)$ with $F(T)=F_0+AT^2$.}}
	\end{center}
\end{table}

\subsection*{B. Fitting of experimental results}

The analysis procedure described in A. has been applied to all data, measured on different materials as well as different devices. At all temperatures the experimental results can are well described by the general Lifshitz-Kosevich formula for longitudinal magneto-transport[Supplementary \equref{sdhformula}], as presented in Supplementary \fig{fig:FTemp}, Supplementary \fig{fig:FTemp-1}, Supplementary \fig{fig:FTemp-2} for Cd$_3$As$_2$, Bi$_2$O$_2$Se and LaRhIn$_5$ respectively. The temperature dependence of oscillation frequencies for different materials/devices are readily included in Fig. 4. 
To further demonstrate the validity of the temperature dependent frequency fitting, a log-log plot of frequency change $-\Delta F$ versus temperature for all three materials is also presented (Supplementary \fig{fig:log}). It is clear that $T^2$ dependence is the only possible and sufficient description of the experimental results, the physical origin of which is described in \s{sec:consistency} in details. {For a further consistency check, we have also performed a standard Lifshitz-Kosevich fit to the temperature depedence of quantum oscillation amplitude which are obtained via FFT analysis at certain field window (Supplementary \fig{fig:mc}). The results demonstrate the consistency in cyclotron mass value obtained from both methods.
	These results clearly demonstrate not only the topological fingerprint, but also the consistency of such analysis.}

\subsection*{C. Comparison to Landau-fan method}
To demonstrate the advantage of our method, we have also performed the widely-used Landau-Fan plot method to both Cd$_3$As$_2$ and Bi$_2$O$_2$Se (Supplementary \fig{fig:Landau}). This method clearly fails to distinguish the topological Cd$_3$As$_2$ and topologically-trivial Bi$_2$O$_2$Se as the linearly-extrapolated residual Landau index is $\approx$ 0.5 for both materials. Such failure is likely resulted from the large SOC of Bi$_2$O$_2$Se which contributes strongly to the oscillation phase despite its topologically trivial nature, as elaborated in previous sections.

\section*{Supplementary Note 4: Distinguishing between 3D Dirac points and  nodal lines in LaRhIn$_5$}\la{sec:pointvslines_larhin}

Here we address if the $7$ T Dirac-type orbit of LaRhIn$_5$  encircles a 3D Dirac point or links with a nodal line, a discussion of interest for the specific physics of this compound. The nodal-line scenario was proposed by Mikitik and Sharlai\cite{La115_Miktik_PRL} based on a speculation that the spin-orbit interaction is weak, but we will instead argue for a Dirac point that is stabilized by the four-fold rotational symmetry of LaRhIn$_5$. Here we demonstrate that the spin-orbit-induced energy splitting of the nodal line is at least comparable in magnitude to the Fermi energy $E_F$ measured from the nodal degeneracy; based on arguments presented in \s{sec:distinguishpointline}, this rules out the nodal-line scenario. 

We have previously determined from our high-temperature oscillation analysis that $|E_F|=|\partial E/\partial \log m_c| \approx 24$ meV, based on the interpretation of a Dirac-type pocket. All that remains is to show that $E_{soc}$ is at least comparable to 24 meV, and indeed our first-principles  estimate for $E_{soc}$ gives a value ranging from $20-100$ meV, as described in \s{sec:bandstructurecal} below. To further support the Dirac-point scenario, we describe a possible $\bk \cdot  \bp$ model in \s{sec:kdotpdirac}.

\subsection*{A. Spin-orbit energy scale and  Dirac-type Fermi surfaces from  band-structure calculations}\la{sec:bandstructurecal}

Band-structure calculations were performed with the goals of estimating $E_{soc}$ and reproducing the experimentally observed Dirac-type pocket. Using the Vienna ab initio Simulation Package (VASP)\cite{PhysRevB.54.11169,PhysRevB.48.13115}, we performed a density-functional calculation within the generalized gradient approximation (GGA), with projector-augmented-wave (PAW) potentials\cite{PhysRevB50.17953,PhysRevB.59.1758} and a 7$\times$ 7 $\times$ 5 $\Gamma$-centered k-point grid for Brillouin zone sampling. Wave functions were expanded in a plane-wave basis up to an energy cutoff of 230 eV. If our first-principles calculation is performed ignoring the spin-orbit interaction, we find a nodal-line degeneracy within the high-symmetry, mirror-invariant planes, as illustrated by red lines in Supplementary \fig{fig:abinitio}(b). While, with the inclusion of spin-orbit coupling (calculated by a second-variational procedure), we show with the blue lines of Supplementary \fig{fig:abinitio}(b) that the nodal-line degeneracy is lifted on a scale $E_{SOC}\approx (20-100)$ meV. 

To check for the robustness of these results against choices of functionals in the calculation, a slightly different calculation was performed in VASP, with the PAW basis sets~\cite{PhysRevB50.17953} combining GGA. The exchange-correlation functional of Perdew, Burke and Ernzerhof (PBE)~\cite{PhysRevLett.77.3865} is added in this case. The cutoff energy for the plane wave expansion was set to 500~eV and a $k$-point mesh of $9\times9\times6$ was used in the bulk calculations. The Irvsp code~\cite{irvsp} was used to obtain the irreducible representations of Bloch states. The obtained band structures with and without spin-orbit coupling (SOC) are shown in Supplementary \fig{fig:abinitio}(c) and (d). Without SOC, the calculation confirms that
LaRhIn$_5$ is a nodal-line metal, e.g., one point on this nodal line is a crossing between two irreducible  representations $\Gamma_2$ and $\Gamma_4$ on the $\Gamma$-M line, as shown by the touching between blue and green energy levels in Supplementary \fig{fig:abinitio}(c).   All band crossing are gapped when SOC is included except the one along the $\Gamma$-Z direction which is protected by C$_{4v}$ symmetry. The irreducible representations $\Gamma_6$ and $\Gamma_7$ are degenerate at a Dirac point on the $\Gamma$-Z line, as shown in Supplementary \fig{fig:abinitio}(d).

While both the comparison of energy scales and symmetry analysis support the Dirac-point scenario, there are deficiencies in our band-structure calculation that prevent full confidence. Foremost is that the calculated oscillation frequencies (of the Brillouin-zone-sized pockets) deviate from the measured frequencies [$F\sim (1-10)kT$] by about $10\%$. This makes it unlikely that the calculation would reproduce fine features of order $7$ T, and indeed  the calculation does not produce 3D Dirac points within a $25$ meV window of the Fermi level -- though it does produce 3D Dirac points at $\approx 420$ meV below the Fermi level, which may or may not be a red herring; this calculated Dirac point is circled in Supplementary \fig{fig:abinitio}(a) and (d). Since all that our argument needs is an order of magnitude estimate for $E_{soc}$, the argument's conclusion should not be overly sensitive to minor deficiencies of the calculational method. It is hoped that future refinements of the band-structure model of LaRhIn$_5$ will settle this matter conclusively.

\subsection*{B. Possible $\bk\cdot\bp$ model for 3D Dirac fermions}\la{sec:kdotpdirac}

To further support the Dirac-point scenario, we performed a $\bk \cdot  \bp$ analysis assuming only that the four independent wave functions (at the hypothesized Dirac crossing) are predominantly Rhodium $d_{z^2}$ and $d_{x^2-y^2}$ orbitals, with each orbital doubled owing to spin. These same wave functions span the degenerate subspace of the calculated Dirac crossing, as described in \s{sec:pointvslines_larhin}.

Postponing the technical derivation below, we first summarize the salient conclusions of our $\bk \cdot  \bp$ analysis:
The linearized  $\bk \cdot  \bp$ Hamiltonian is simply a direct sum of two Weyl Hamiltonians with opposite chirality. Chirality being a conserved quantity is an emergent property of \textit{small}, rotation-invariant Dirac FS's, and is not generally expected for Cd$_3$As$_2$/Na$_3$Bi. One implication of conserved chirality is that for a field parallel to the rotation axis $c$, $\lambda_1^{\uparrow}=\lambda_1^{\downarrow}=\pi$, with ${\uparrow/\downarrow}$ viewed as a chirality index. The $\pi$ phase is completely attributed to a Berry phase; the contribution by the Zeeman interaction vanishes owing to spin being locked in the plane orthogonal to the rotation axis\cite{phiB_Aris_PRX}. This linearized model gives a plausible account for the measured $\lambda_1^{\uparrow}=(0.98\pm 0.01)\pi$; the small deviation from $\pi$ is plausibly attributed to higher-order corrections in the $\bk\cdot\bp$ Hamiltonian. By adding a Hamiltonian term that is quadratic in $k_z$ and proportional to the identity matrix, the Fermi surface can be transformed to have a `neck'  with a minimal cross-section at $k_z=0$ -- this would account for the Lifshitz-Kosevich phase $\phi_{LK}=+\pi/4$ for $B//c//k_z$ parallel to the rotational axis. (A similar-looking `neck' Fermi surface is illustrated in Fig.\ 1(b) of \ocite{La115_Miktik_PRL} and derived from a different $\bk \cdot  \bp$ model.)

Our $\bk\cdot \bp$ model reproduces all salient features of our data for $B//c$, but does not, however, explain the data at various tilt angles of $B$. As illustrated in Supplementary \fig{fig:anglepolar}(b), a quantum-oscillation frequency $F$ between 7 to 8 T persists over a 90$^{\circ}$ tilt in the $a-c$ crystallographic plane, suggesting a nearly isotropic Fermi surface. However, the observation of $\phi_{LK}=+\pi/4$ caution against the simplest interpretation of a closed isotropic pocket as it corresponds to a minimum instead of maximum orbit. We have not been able to infer a completely explanatory model based on the present data. However it is worth remarking that the persistence of $F\approx 7$ T over 90$^{\circ}$ is even harder to explain in the nodal-line scenario: the Fermi surface of a nodal-line semimetal is highly anisotropic and is known (in some cases\cite{pezzini2018unconventional}) to exhibit magnetic breakdown depending on the field orientation.

For a detailed $\bk\cdot\bp$ analysis, let us define the tensor products $\{\sigma_i\otimes \tau_j\}$ of two Pauli matrices as a basis for Hermitian operators in the four-dimensional vector space, with the correspondence
\begin{equation}
\tz=+1 \leftrightarrow d_{z^2}, \;\; \tz=-1 \leftrightarrow d_{x^2-y^2},\;\;\sz=+1\leftrightarrow \uparrow,\;\; \sz=-1\leftrightarrow \downarrow,
\end{equation}
with $\uparrow/\downarrow$ referring (in this appendix) to spin up and down along the rotation-invariant $z$ axis.  The group of the wavevector at the Dirac point includes the point group $C_{4v}$, which is generated by a four-fold rotation $\rot_{4z}$ and a reflection $\mir_x$ that inverts $x\rightarrow -x$. In addition, the groups of all wavevectors have the symmetry $T\inv$, which is the composition of time reversal with spatial inversion. In the four-dimensional vector space, the matrix representations of the symmetry generators are
\begin{equation}
T\inv = i\sy K, \as \rot_{4z}=e^{-i\sz \pi/4}\tz, \as M_x=-i\sx. 
\end{equation}
The first symmetry ensures that the $\bk\cdot \bp$ Hamiltonian $H(\bk)$ can only contain five anticommuting matrices: 
\begin{equation}
\sx\ty,\;\sy\ty,\;\sz\ty,\;\tx,\;\tz, 
\end{equation}
ignoring the identity matrix. Let us keep only terms in $H(\bk)$ which are independent of or linear in $\bk$. The four-fold symmetry ensures that 
\begin{equation} H'(\bk)=v_{\parallel}(k_y\sx+k_x\sy)\ty+v'_{\parallel}(k_x\sx-k_y\sy)\ty  + (v_{\perp}k_z+\var_0)\tz. 
\end{equation}
The reflection symmetry allows us to set $v'_{\parallel}=0$. Redefining the zero of $k_z$ to lie at the Dirac point, we are left with
\begin{equation} H(\bk)=v_{\parallel}(k_y\sx+k_x\sy)\ty +v_{\perp}k_z\tz. 
\end{equation}
By a unitary transformation, the Hamiltonian is expressible as a direct sum of two (3+1)D Weyl Hamiltonian with opposite chirality: 
\begin{equation} U=\tf1{\sqrt{2}}(I+i\sz\tx), \as UH(\bk)U^{-1} = v_{\parallel}(k_y\sx+k_x\sy)\ty +v_{\perp}k_z\sz\ty:= H'(\bk),
\end{equation}
with the chirality operator being $\ty$. The following identity is useful to derive the above expression:
\begin{equation}
\begin{aligned}
U=\tf1{\sqrt{2}}(I+iA), \as A^2=I,\as UBU^{-1}=\begin{cases}B, &[A,B]=0 \\ \f{i}{2}[A,B], & \{A,B\}=0. \end{cases}
\end{aligned} 
\end{equation}
with $[A,B]$ the commutator and $\{A,B\}$ the anticommutator. In the new basis, the Hamiltonian has the  constraint:
\begin{equation} \mir_z H'(\bk)\mir_z=H'(k_x,k_y,-k_z), \as \mir_z=\sz\tz,\la{emergentref}
\end{equation}
even though the reflection $z\rightarrow -z$ is not in the little group of the wavevector; such an `emergent' symmetry will be spoiled by corrections to $H(\bk)$ which are higher order in $\bk$. \\

For simplicity, let us consider applying a field parallel to the rotation axis. Owing to the emergent reflection symmetry of Supplementary \equref{emergentref}, the extremal orbit lies at $k_z=0$, where $H'(k_x,k_y,0)=v_{\parallel}(k_y\sx+k_x\sy)\ty$ simplifies to a direct sum of two (2+1)D Dirac Hamiltonians. Because chirality is conserved under parallel transport (in this linearized approximation), the Berry phase can be calculated separately for each Dirac fermion and simply equals $\pi$. The orbital magnetic moment matrix can be expressed in terms of inter-band matrix elements of the position operator [see Eq.\ (65) in \ocite{100page}], however such matrix elements vanish between two states with distinct chirality. We then apply that the single-band orbital moment for a (2+1)D Dirac fermion vanishes\cite{phiB_Aris_PRX}. Finally, the spin operator $S_z=(\hbar/2)\sigma_z$ in the original basis is still represented as $(\hbar/2)\sigma_z$ in the new basis. Since $[S_z,\tau_y]=0$, the Zeeman spin interaction also does not couple states with distinct chirality; since the spin of each Dirac fermion lies in the plane orthogonal to the rotational axis (and also orthogonal to the field), the Zeeman spin interaction has a vanishing effect. In combination, the quantization offset $\lambda$ simply equals a Berry phase of $\pi$, for both chiralities of Weyl fermions.

\section*{Supplementary Note 5: Further characterization of LaRhIn$_5$}

\subsection*{A. Angular dependence of Shubnikov-de Haas oscillation frequencies}

Fast-Fourier-Transformation (FFT) is performed to analyze the Shubnikov-de Haas (SdH) oscillations of the high frequencies belonging to the large, trivial Fermi surfaces. In contrast, the Dirac pocket was analyzed by directly fitting Supplementary \equref{truefit} as appropriate for such small frequencies (see SIV). The high-frequency SdH oscillations were measured at $T = 2$~K with current applied along $c$-axis and field rotates from $c$ to $a$-axis [Supplementary \fig{fig:FFTangle}(a)]. At each angle the results are shifted and normalized according to the largest peak, the square root of oscillation amplitudes are taken for visualizing the small details. The frequencies corresponding to the larger Fermi surfaces $\alpha_1$, $\beta_1$ and $\beta_2$ are denoted by the dashed lines. The angular dependence of oscillation frequencies [Supplementary \fig{fig:FFTangle}(b)] are obtained by identifying the positions of major peaks in the FFT spectrum. The angular dependence of these high-frequency oscillations corresponding to the large FS cross-sections [solid symbols in Supplementary \fig{fig:FFTangle}(b)] agrees well with previous dHvA results\cite{LaCe115_dHvA_jpsj} indicated by dashed lines.

\subsection*{B. Dingle analysis}

Clearly obserable quantum oscillations observed at $T = 2$~K with current applied along $c$-axis and field applied along $a$-axis [Supplementary \fig{fig:Dingle}(a)] makes it possible to determine the quantum mean free path value. FFT analysis is performed within different field windows [Supplementary \fig{fig:Dingle}(b)] for the quantum mean free path analysis. The enhanced peak value at larger magnetic field represents the increasing ratio between Landau level spacing and broadening due to scattering, as expected for the Dingle damping effect. The quantum mean free paths of different extremal orbits are therefore calculated by fitting the field dependence of oscillation amplitudes[Supplementary \fig{fig:Dingle}(c)] in consideration of both thermal and Dingle damping terms. The results yield long quantum mean free paths ($300~$to$~900~nm$) as expected for the clean LaRhIn$_5$ sample.

\newpage

\begin{suppfigure}
	\includegraphics[width=0.6\columnwidth]{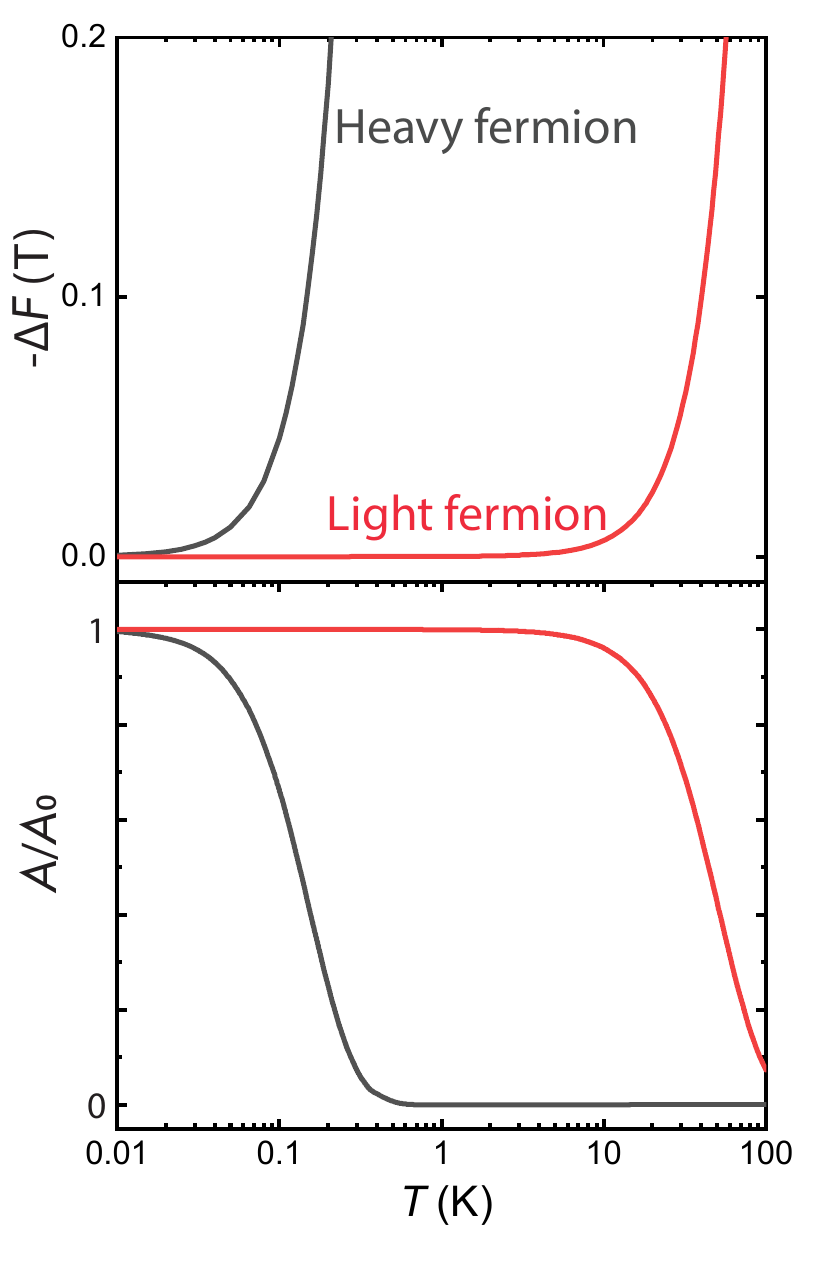}
	\caption{ Illustration of frequency change vesus smearing of oscillation amplitude due to increasing temperature. Here zero-temperature oscillation frequency ($F_0$) is chosen to be 50 T for both cases, while cyclotron mass is 10$m_e$ and 0.03 $m_e$ for heavy and light fermion respectively.}\la{fig:mass}
\end{suppfigure}

\begin{suppfigure}
	\includegraphics[width=0.95\columnwidth]{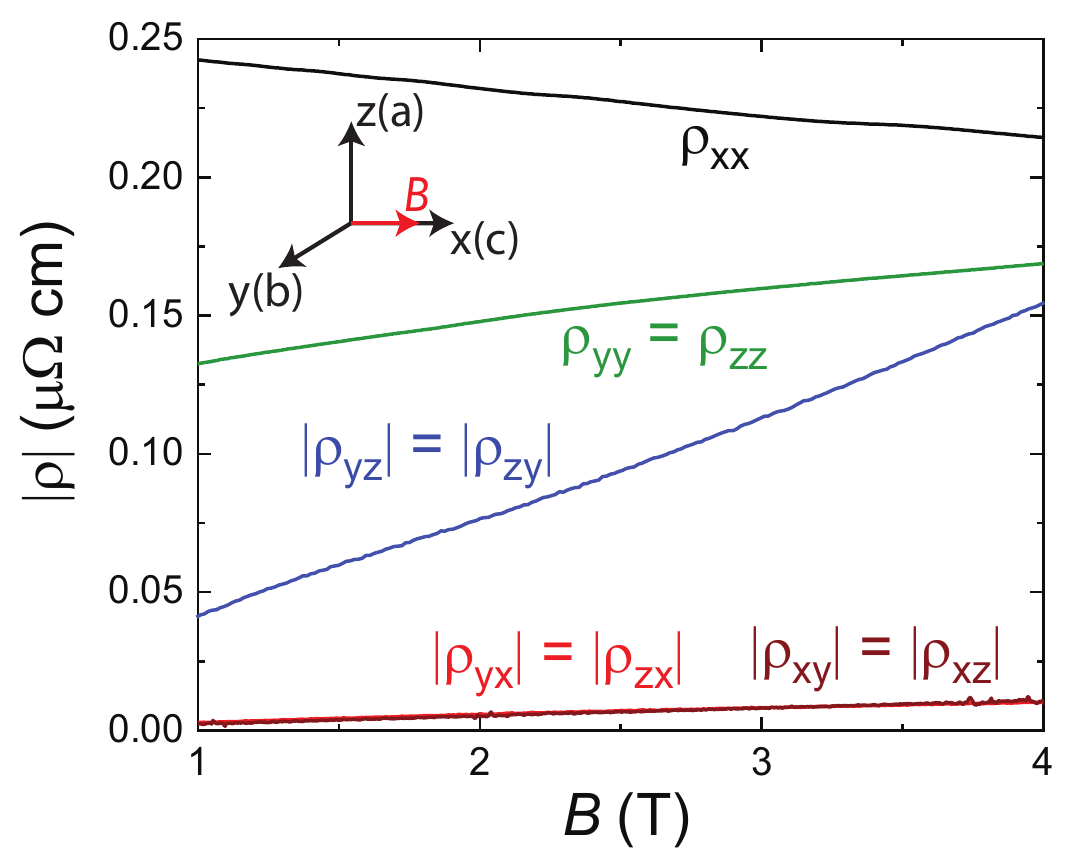}
	\caption{All components of resistivity tensor of LaRhIn$_5$ measured with magnetic field applied along $x(c)$-axis. The planar Hall components ($\rho_{xy},\rho_{xz},\rho_{yx},\rho_{zx}$) is much smaller than the Hall resistivity($\rho_{yz},\rho_{zy}$) and magnetoresistivity($\rho_{xx},\rho_{yy},\rho_{zz}$) components.}\la{fig:Rtensor}
\end{suppfigure}

\clearpage

\begin{suppfigure}
	\includegraphics[width=0.95\columnwidth]{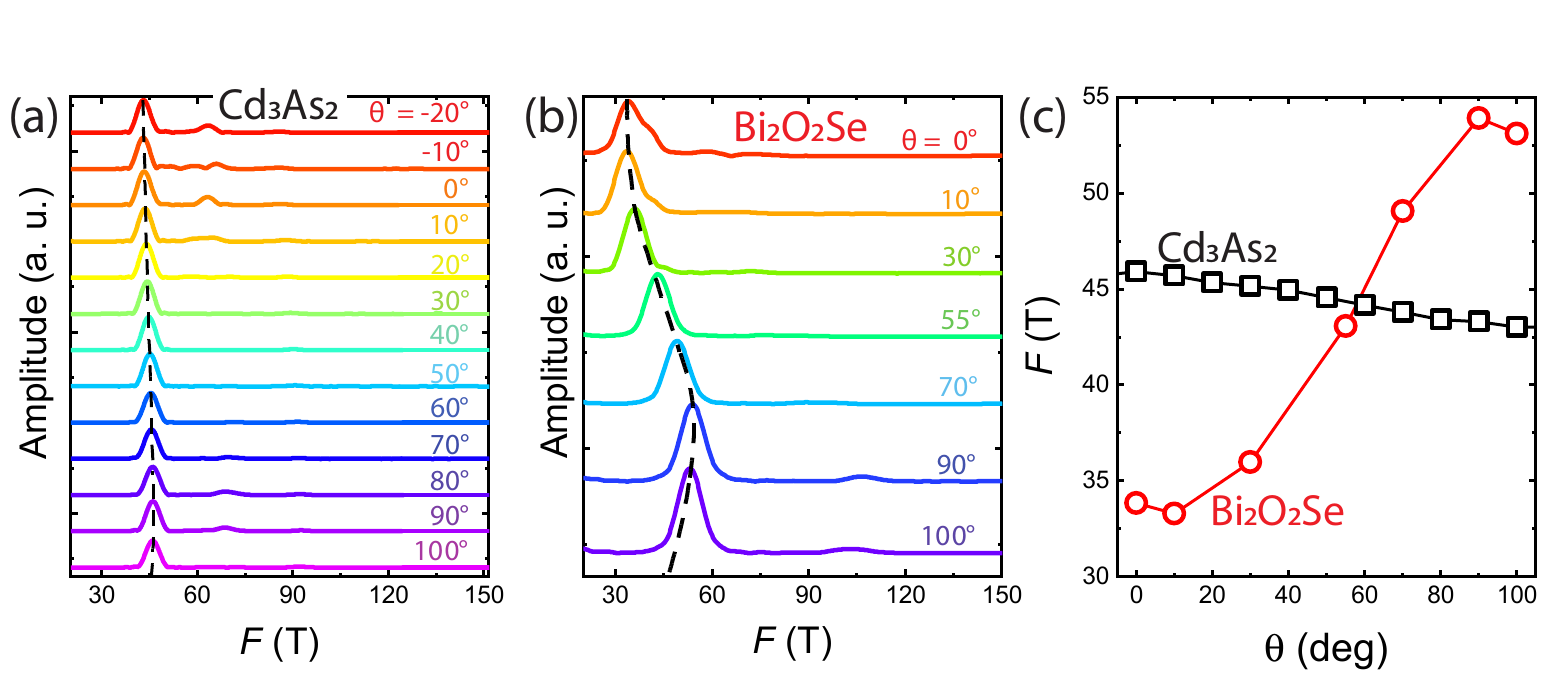}
	\caption{The Fast-Fourier-Transformation (FFT) spectrum at different angles for Cd$_3$As$_2$ and Bi$_2$O$_2$Se respectively. (a) For Cd$_3$As$_2$, $\theta$ stands for the angle between applied field and [111]-axis. (b) For Bi$_2$O$_2$Se, $\theta = 0^{\circ}$ is defined as field applied along c-axis.}\la{fig:anglepolar-2}
\end{suppfigure}

\begin{suppfigure}
	\includegraphics[width=0.95\columnwidth]{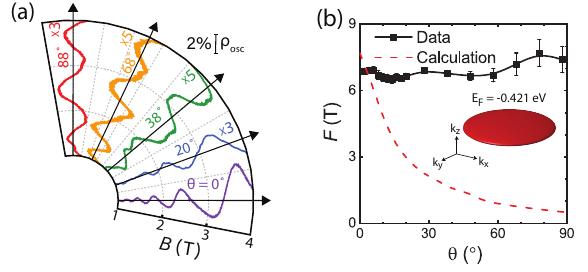}
	\caption{Angle-dependent oscillation frequency corresponds to the small Dirac pocket in LaRhIn$_5$. (a) Fan plot of angle-dependent quantum oscillations ($F\approx$ 7 T) at $T = 2$ K. $\theta$ is the angle between the magnetic field and the $c$-axis. (b) Angular dependence of measured oscillation frequency (black) deviates from the ab-initio calculation (red). The error bar is defined by the standard error of the fitting parameters generated by the fitting of quantum oscillations presented in (a). The calculated Fermi surface of the Dirac node at $E_F$ = -0.421 eV is illustrated in the inset.}\la{fig:anglepolar}
\end{suppfigure}

\clearpage

\begin{suppfigure}
	\includegraphics[width=0.95\columnwidth]{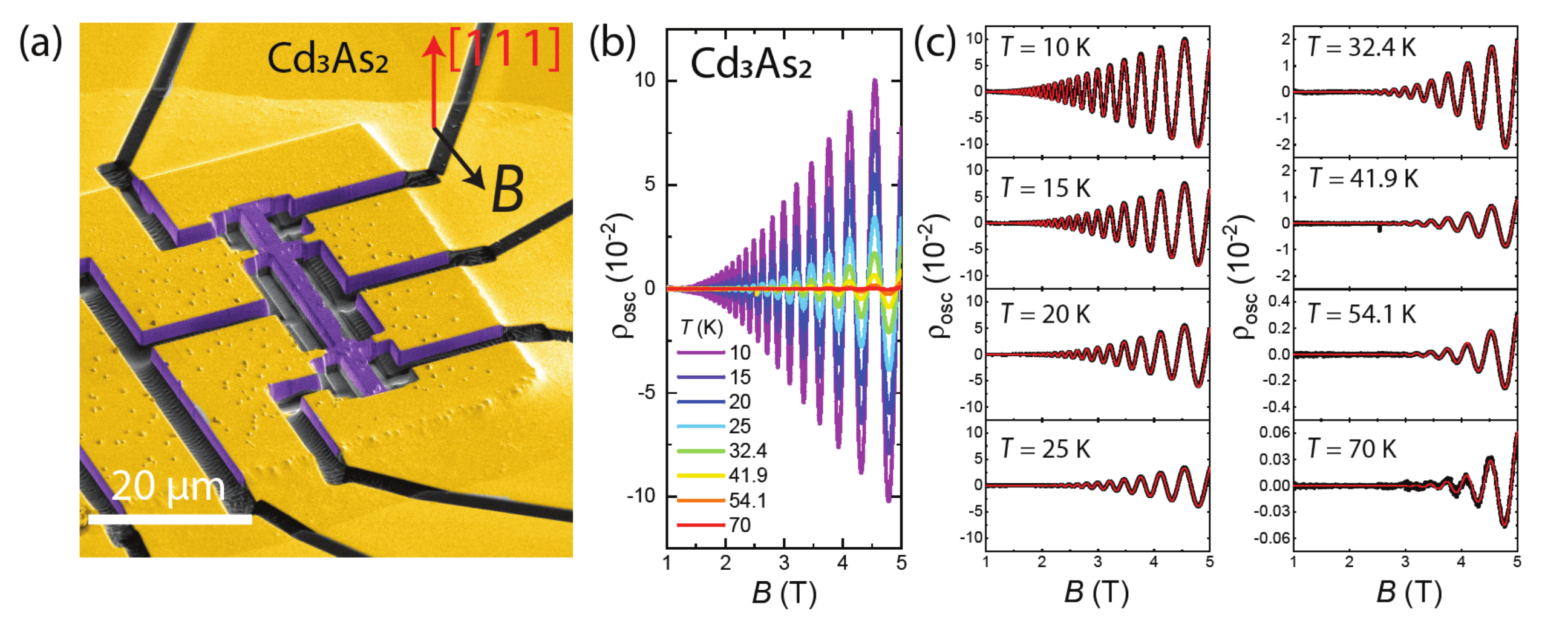}
	\caption{Quantum oscillation measurements of Cd$_3$As$_2$. (a) Scanning Electron beam micrograph of a FIB prepared Cd$_3$As$_2$ microstructure. The crystalline pathway (purple) is electrically connected via gold top contacts (yellow). The magnetic field is applied perpendicular to the out-of-plane [111] axis. (b) temperature dependent quantum oscillations, (c) comparison of the experimental results (black curve) to the LK fit (red) at all temperatures.}\la{fig:FTemp}
\end{suppfigure}

\begin{suppfigure}
	\includegraphics[width=0.95\columnwidth]{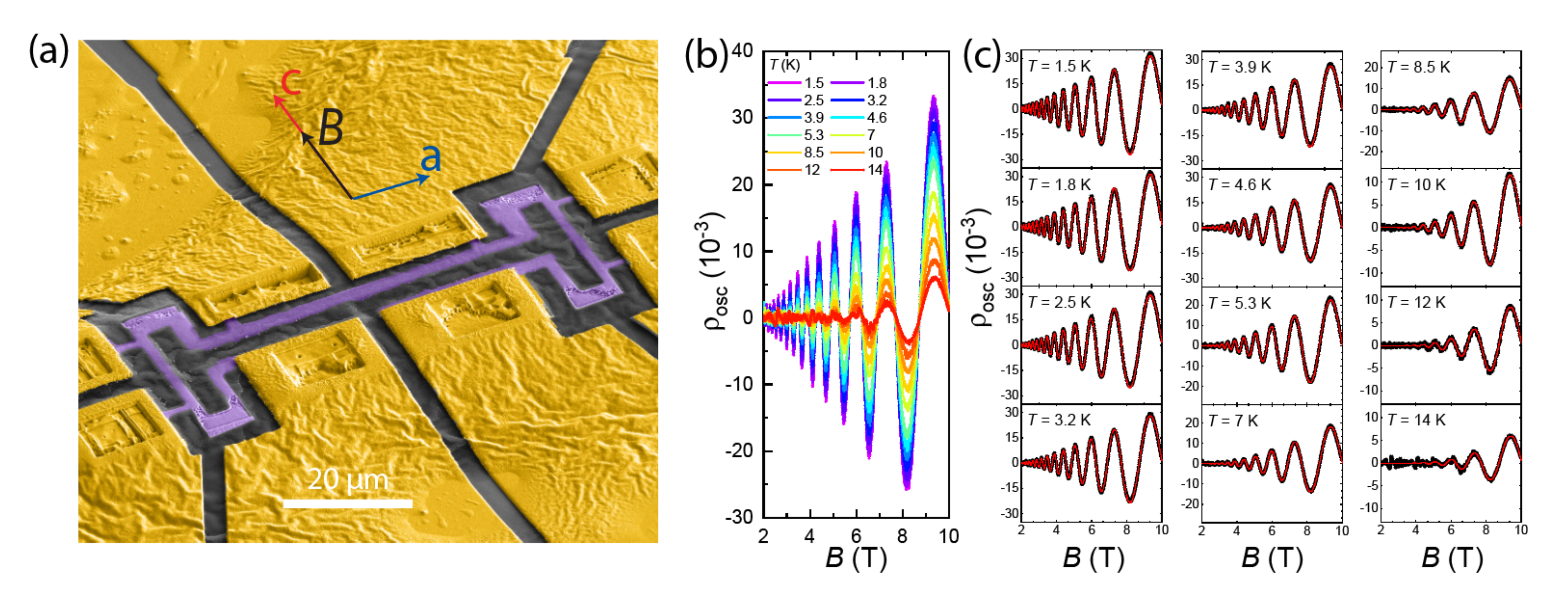}
	\caption{Quantum oscillation measurements of Bi$_2$O$_2$Se. (a) Scanning Electron beam micrographs of the Bi$_2$O$_2$Se device. The magnetic field is applied along $c$-axis.(b) temperature dependent quantum oscillations, (c) comparison of the experimental results (black curve) to the LK fit (red) at all temperatures.}\la{fig:FTemp-1}
\end{suppfigure}

\begin{suppfigure}
	\includegraphics[width=0.95\columnwidth]{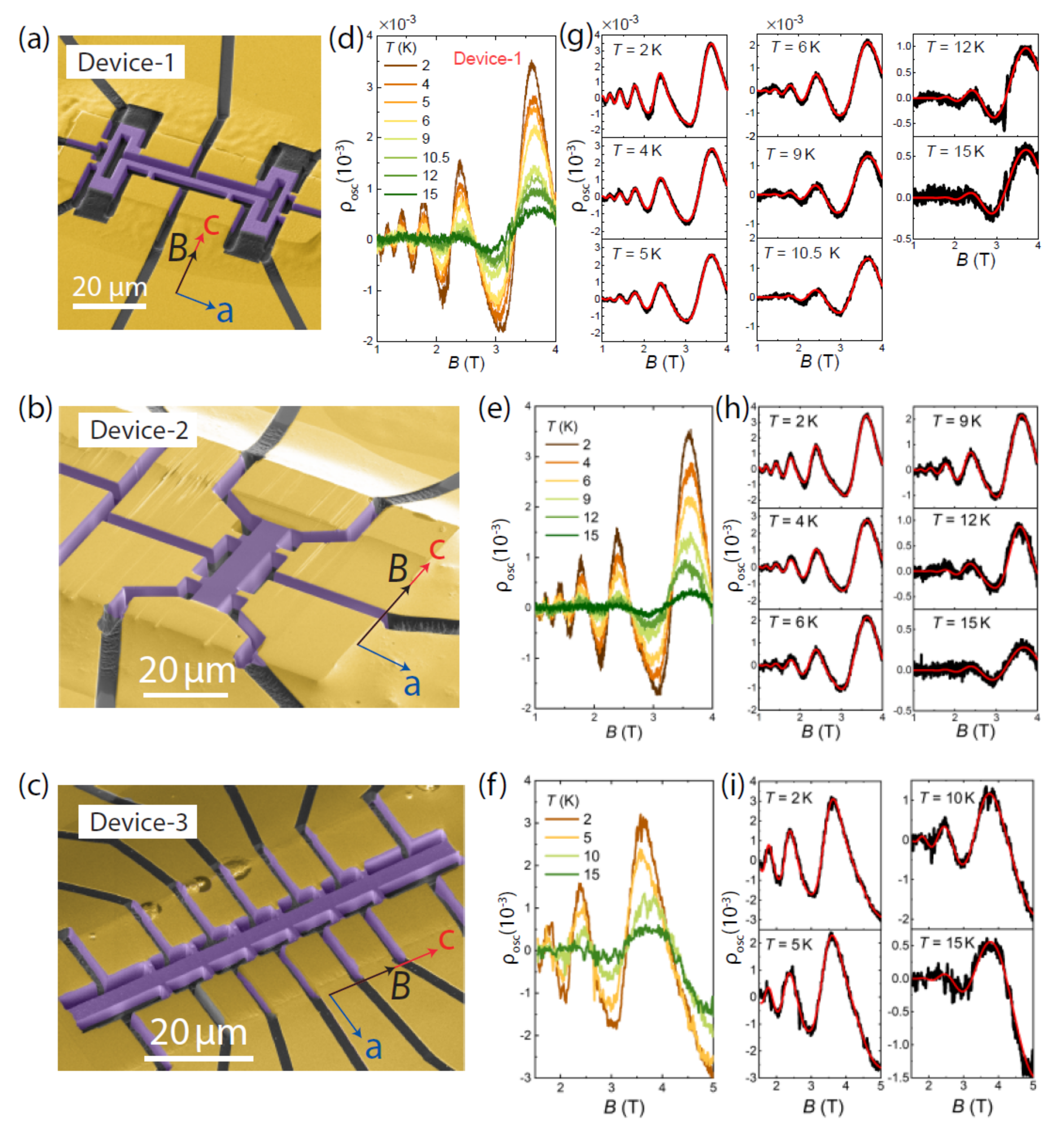}
	\caption{Quantum oscillation measurements of LaRhIn$_5$. (a), (b) and (c) display the Scanning Electron beam micrographs of devices 1-3 of LaRhIn$_5$ respectively. (d), (e) and (f) present the temperature dependent quantum oscillations, while (g), (h) and (i) show the fits (red curve) to the experimental results (black) at all temperatures.}\la{fig:FTemp-2}
\end{suppfigure}

\begin{suppfigure}
	\includegraphics[width=0.95\columnwidth]{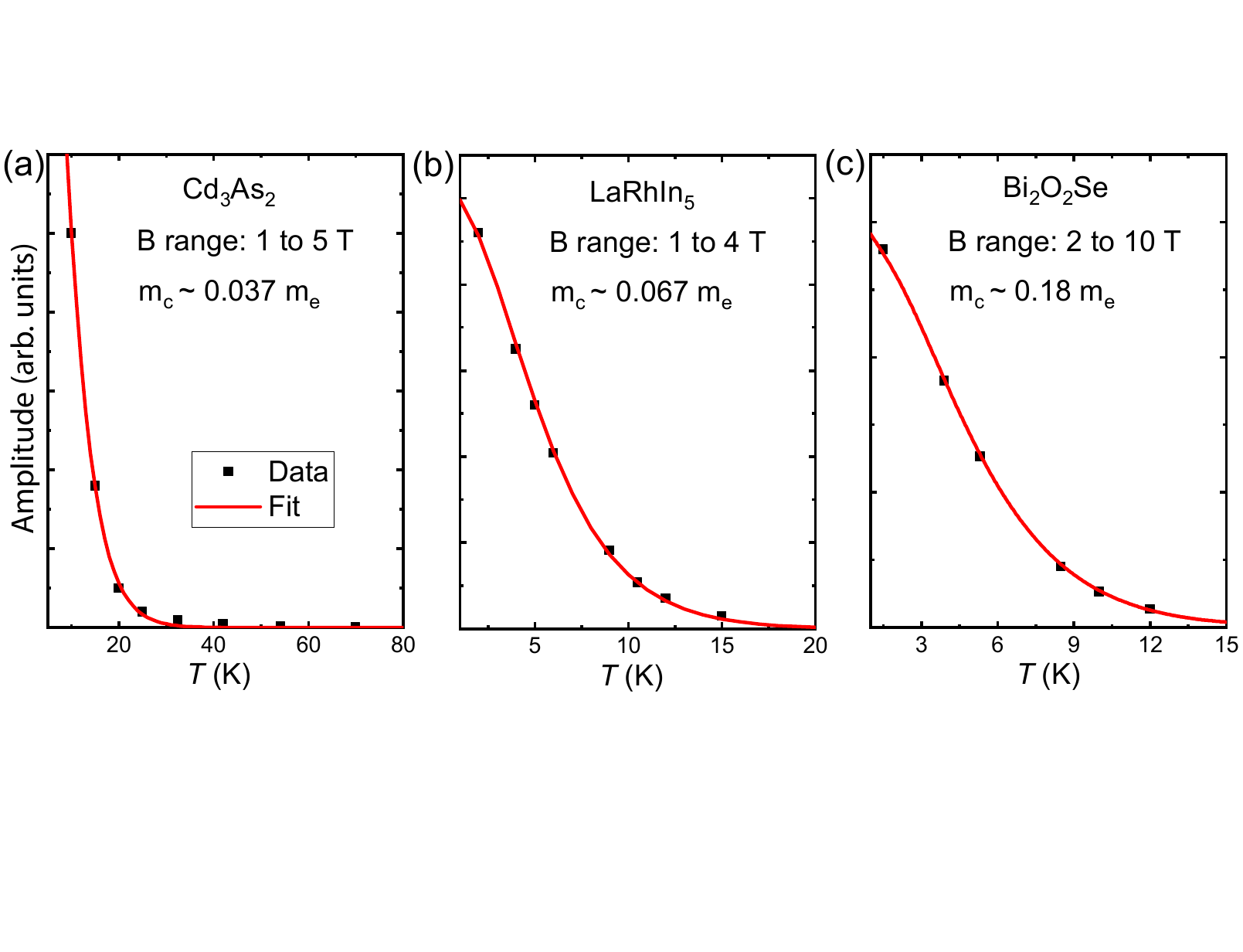}
	\caption{{Temperature dependence of oscillation amplitude for Cd$_3$As$_2$ (a), LaRhIn$_5$ (b) and Bi$_2$O$_2$Se (c). The fitting yields consistent value of cyclotron mass with the direct fitting described in \s{sec:fit} }}\la{fig:mc}
\end{suppfigure}

\begin{suppfigure}
	\includegraphics[width=0.95\columnwidth]{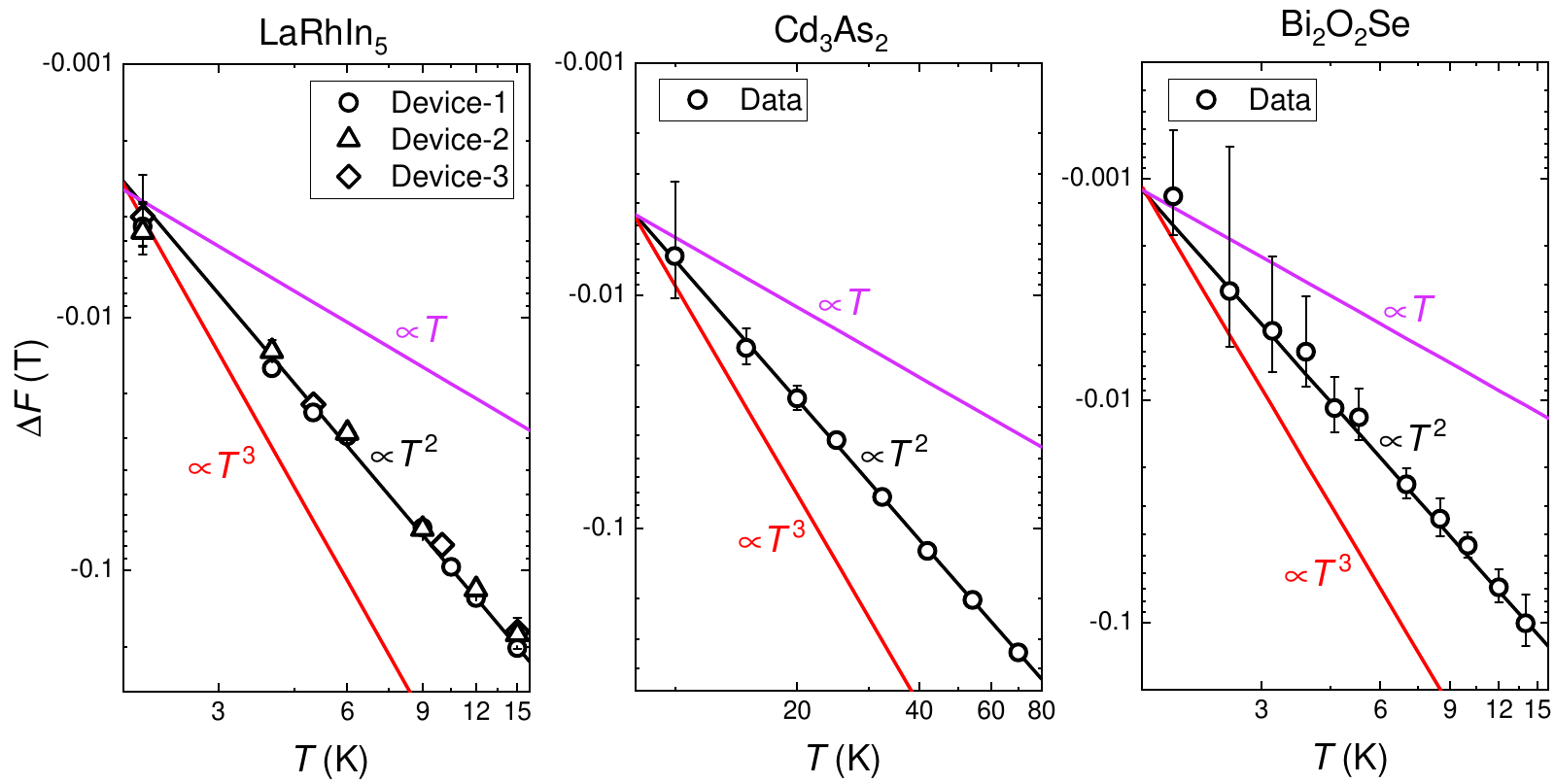}
	\caption{Log-log plot of temperature dependence of oscillation frequency change $\Delta F$ for LaRhIn$_5$, Cd$_3$As$_2$ and Bi$_2$O$_2$Se. The error bar is determined by the standard error of the fitting parameters generated by the non-linear regression fitting procedure. It is clear that $T^2$ law is the only meaningful description of our data.}\la{fig:log}
\end{suppfigure}

\begin{suppfigure}
	\includegraphics[width=0.95\columnwidth]{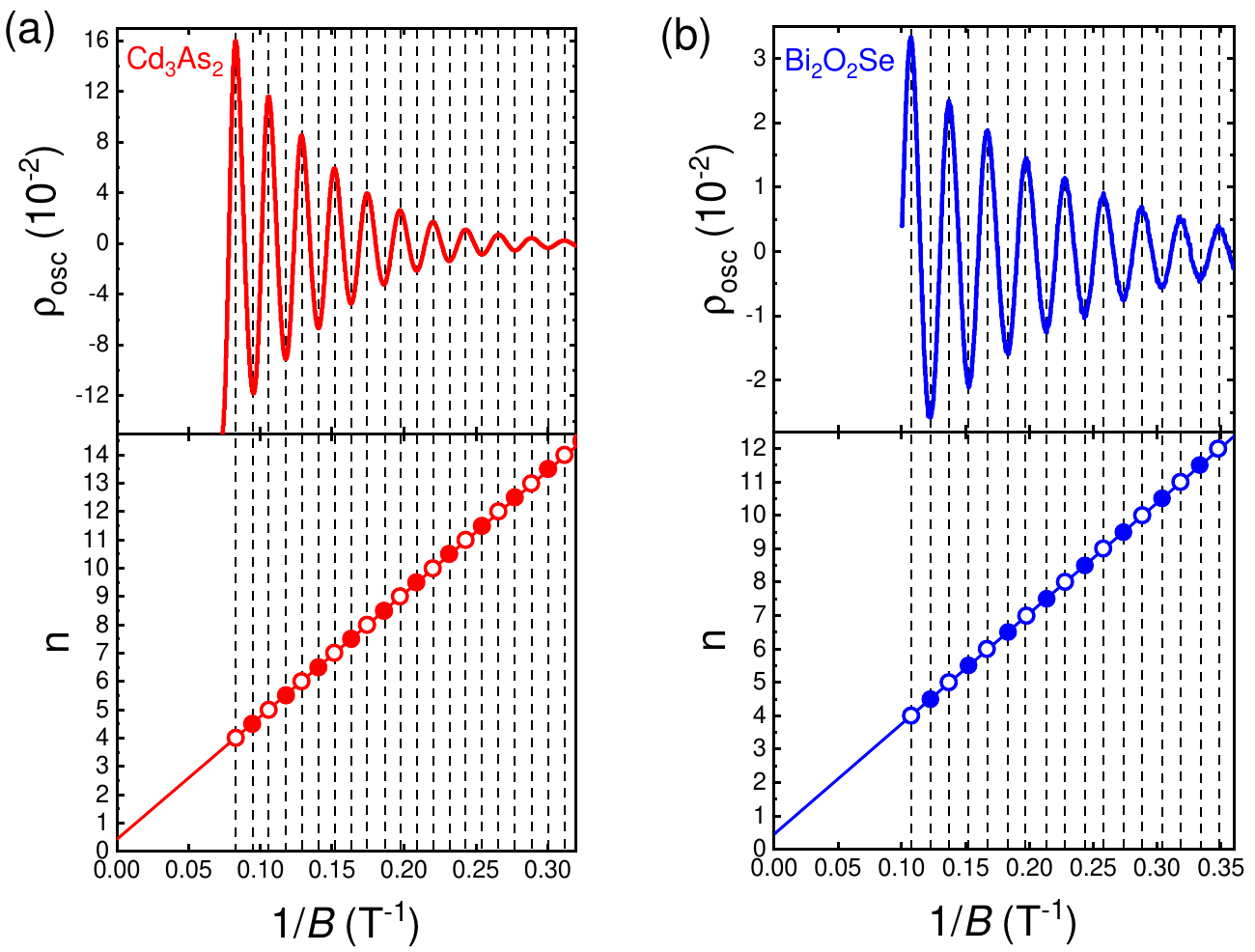}
	\caption{Landau-fan plot of Cd$_3$As$_2$ (a) and Bi$_2$O$_2$Se (b) respectively. For SdH oscillations measured with the longitudinal configuration, the peaks of the oscillations correspond to the integer number(open symbol) of Landau index, while the valleys correspond to the half-integer (solid symbol) ones. The analysis for both materials demonstrates a residual Landau index close to 0.5, which may be incorrectly interpreted as both materials being "topological", while Bi$_2$O$_2$Se is clearly topologically-trivial. }\la{fig:Landau}
\end{suppfigure}

\begin{suppfigure}
	\includegraphics[width=0.95\columnwidth]{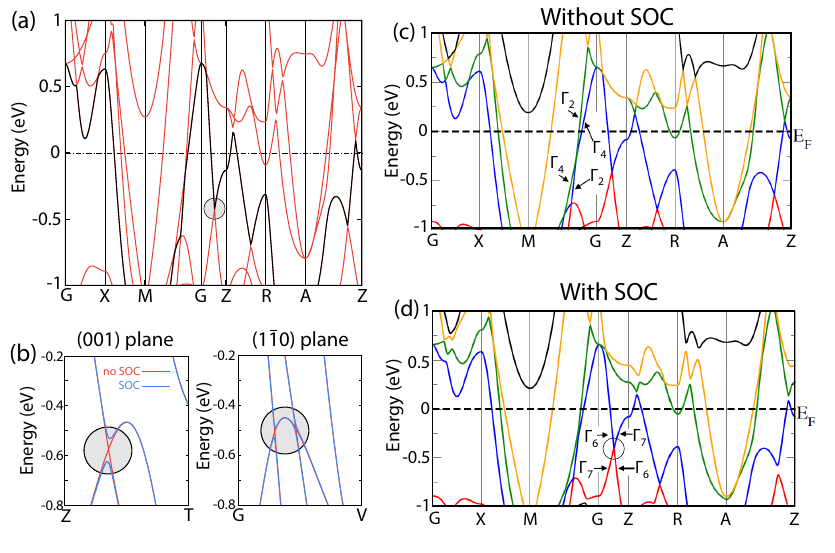}
	\caption{Band structure calculations for LaRhIn$_5$. (a) Ab-initio-calculated band structure of LaRhIn$_5$, with a 3D Dirac point encircled. (b) illustrates the SOC-induced energy gaps at two cross-sections of the nodal line. (c) and (d) compares band structure of LaRhIn$_5$ with and without spin-orbit coupling calculated with the consideration of PBE. $\Gamma_2$ and $\Gamma_4$ in panel (c) are two irreducible representations of little point group C$_{2v}$.  $\Gamma_6$ and $\Gamma_7$ in panel (d) are two irreducible representations of double point group C$_{4v}$\cite{altmann1994}.}\la{fig:abinitio}
\end{suppfigure}

\begin{suppfigure}
	\includegraphics[width=0.95\columnwidth]{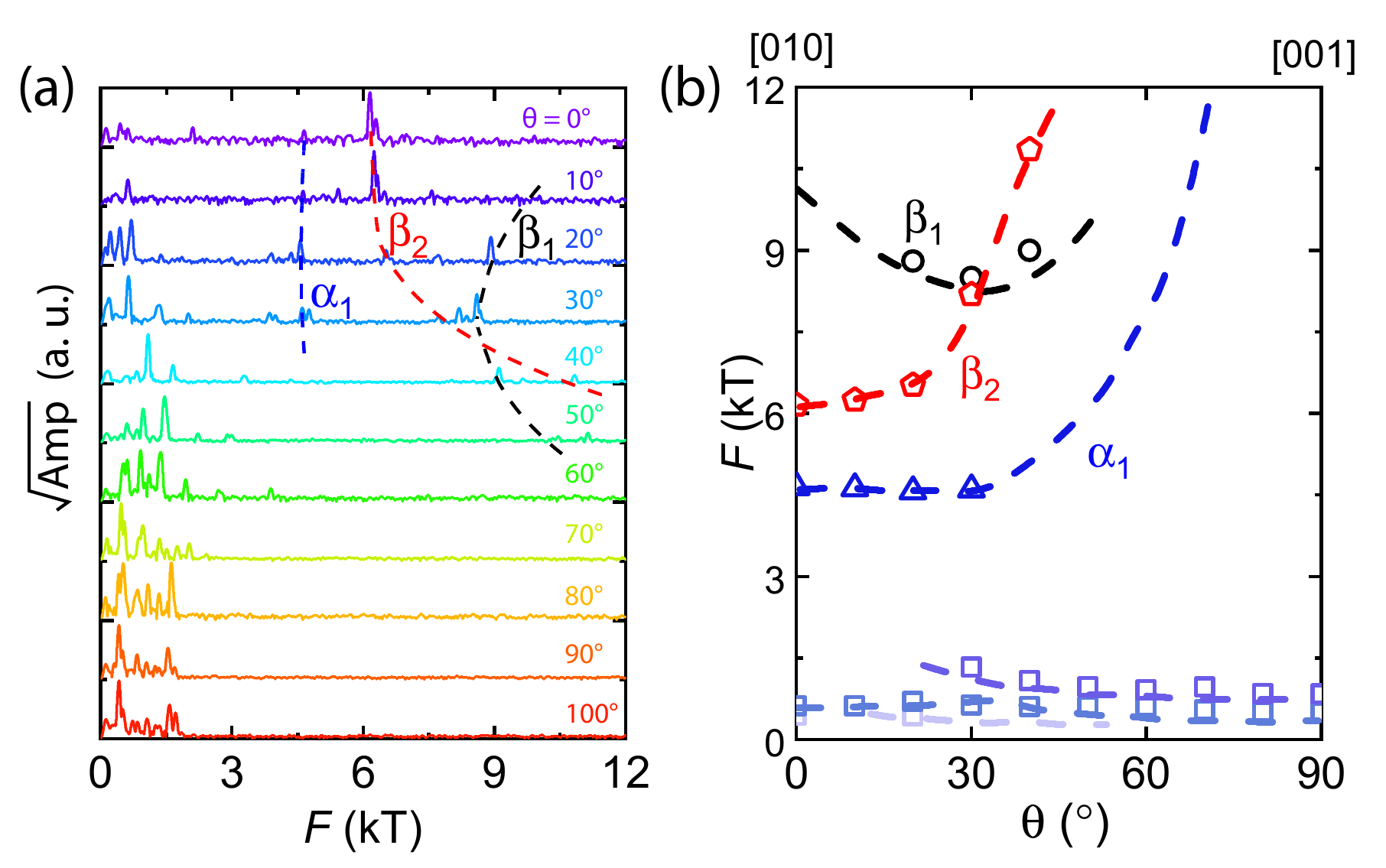}
	\caption{Large-orbit quantum oscillations in LaRhIn$_5$. (a) Angle-dependent FFT spectrum for SdH oscillations with a field window of 8 to 14 T. The angle $\theta$ stands for the angle between magnetic field and $c$-axis. (b) Angular dependence of oscillation frequencies. Here $\theta$ is the angle between the applied magnetic field and the $c$ axis. The solid symbols represent our results, while the results digitized from ref.\cite{LaCe115_dHvA_jpsj} are displayed as the dashed lines. These results indicate that the high crystalline quality of the bulk crystals was not impacted by the microfabrication procedure.}\la{fig:FFTangle}
\end{suppfigure}

\begin{suppfigure}
	\includegraphics[width=0.95\columnwidth]{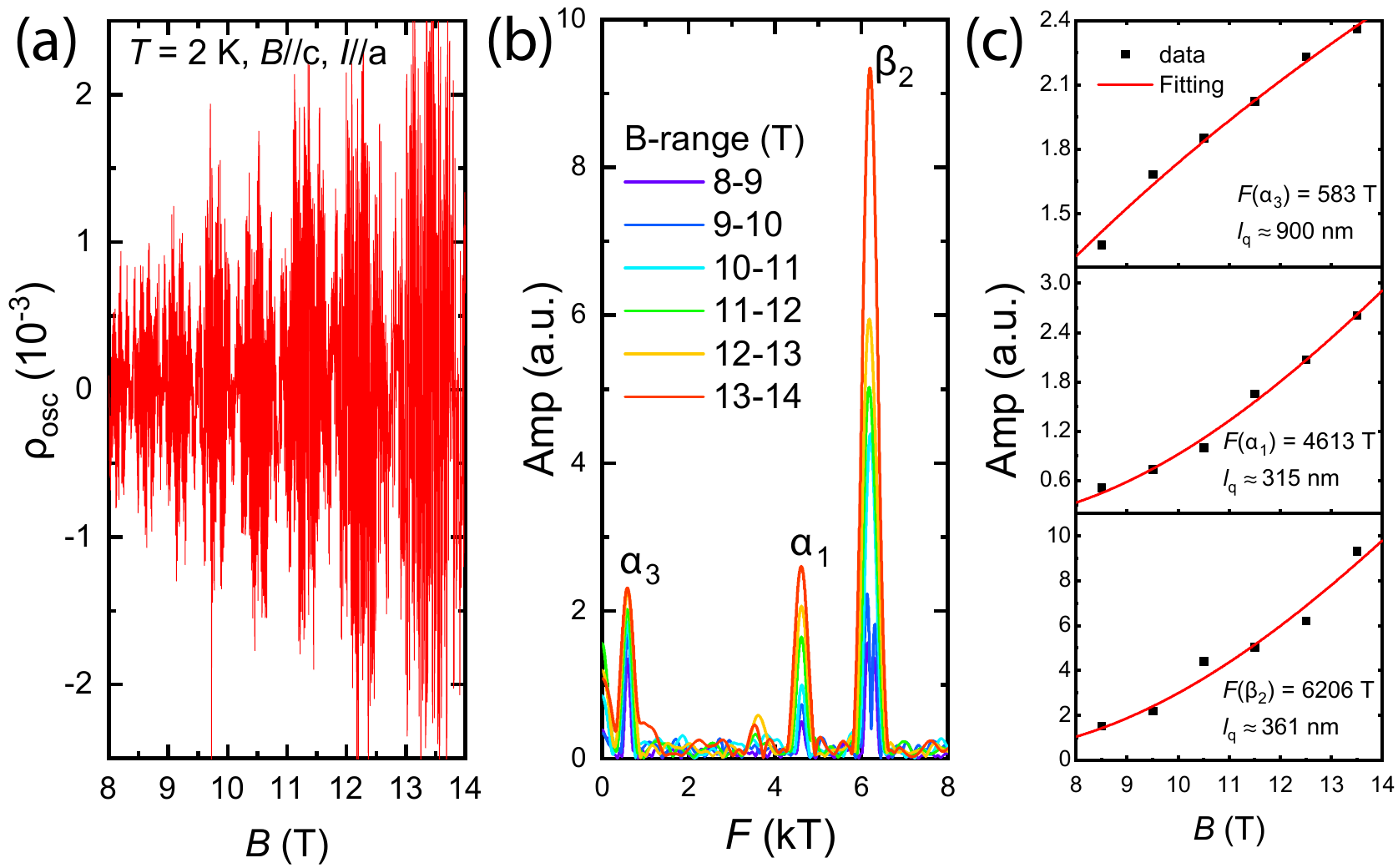}
	\caption{Quantum mean free path analysis in LaRhIn$_5$. (a) Observation of clear high frequency SdH oscillations at $T$ = 2 K with field and current applied along the $c$-axis. The observation of these high frequencies requires magnetic field in excess of the quantum limit of the small pocket, which hence is not observable here. (b) FFT spectrum at $T = 2~$K for different field windows. Three major peaks, which corresponds to $\alpha_3$, $\alpha_1$ and $\beta_2$ orbits are clearly observed. (c) The value of quantum mean free path ($l_q$) is obtained by Dingle analysis. The results suggest long quantum mean free paths.}\la{fig:Dingle}
\end{suppfigure}

\newpage


